\documentclass[12pt]{iopart}

\usepackage{iopams}  
\usepackage{color}
\usepackage{bm}
\usepackage{graphicx}
\usepackage{epstopdf}
\usepackage{booktabs}
\usepackage{esint}
\usepackage{tensor}
\usepackage{soul}

\newcommand*{\rttensor}[1]{\overline{\overline{#1}}}

\begin{document}
\PSST

\title[]{\textcolor{black}{Axial-radial plasma transport and performance of a plasma thruster magnetic nozzle under Bohm’s anomalous diffusion scaling}}

\author{Shaun Andrews$^1$, Raoul Andriulli$^1$, Nabil Souhair$^{1,2}$, Mirko Magarotto$^3$ \& Fabrizio Ponti$^1$}

\address{$^1$ Alma Propulsion Laboratory, Department of Industrial Engineering, University of Bologna, Forli 47122, Italy}
\address{$^2$ Lerma Laboratory Aerospace and Automotive Engineering School, International University of Rabat, Shore Rocade, Rabat 11103, Morocco}
\address{$^2$ Department of Industrial Engineering, University of Padova, Padova 35131, Italy}
\begin{indented}
\item[]March 2025
\end{indented}

\begin{abstract}
\textcolor{black}{Magnetic nozzles (MN) are known to be subject to anomalous non-collisional diffusion mechanisms driven by instabilities and wave-particle interactions. This study therefore employs a fully kinetic axial-radial particle-in-cell (PIC) model to examine the impact of this anomalous diffusion on plasma transport and the propulsive performance of MNs typical of low-power cathode-less radio-frequency (RF) plasma thrusters. A Bohm-type anomalous collisionality scaling ($\nu_{an}=\alpha_{an}\omega_{ce}$) is implemented to simulations of the 150 W-class REGULUS-150-Xe thruster, evaluating both low-power (30 W) and high-power (150 W) operating conditions. The impact on azimuthal electron current formation is assessed, as well as its subsequent effect on thrust generation, momentum and power balance, and overall propulsive efficiency.} A critical value of the Bohm coefficient was found to exist, where the MN expansion transitions from a well-collimated to an under-collimated state and electron transport shifts from being dominated by magnetic advection to being dominated by cross-field diffusion. This critical transition was found to occur within a narrow interval between $\alpha_{an}$=1/128 and 1/64. Beyond this threshold, it is found that the enhanced cross-field transport of electrons inhibits the formation of the typical MN potential barrier, reducing the radial confinement. The downstream potential drop is reduced by up to 15\%. Diamagnetic electron current is diminished in the absence of steep pressure gradients and the $E\times B$ current becomes purely paramagnetic. The MN efficiency is cut from circa 0.5 to 0.2 due to loss of electron thermal energy conversion and increased plume divergence. At the Bohm limit of $\alpha_{an}=1/16$, agreement to experimental thrust profiles of $<20\%$ is achieved in contrast to 48\% overestimation at high-power in the classical case.
\end{abstract}

%
\vspace{2pc}
%
\submitto{\PSST}
%
\maketitle
%
%

\section{Introduction}

The magnetically enhanced plasma thruster (MEPT) is a simple low-cost electric propulsion (EP) concept, where thrust is generated by expanding an electromagnetically (EM) heated plasma through a magnetic nozzle (MN) \cite{b:magarotto2020,b:ahedo2010}. Propellant injected into a dielectric source tube is ionised and heated by a MHz range radio-frequency (RF) or GHz range microwave antenna \cite{b:magarotto2020_2}. The resulting quasi-neutral current-free plasma is then guided by a magneto-static field (in the 100-1000~G range) generated by a set of solenoids or permanent magnets \cite{b:takahashi2019helicon}. Inside the source, the magnetic field confines the plasma and guides propagation of the EM waves. Outside, the discharge plasma is accelerated by diamagnetic currents interacting with the diverging magnetic field forming the MN \cite{b:Hu2021}. Low-power ($<$250~W) devices are typically mesothermal (electron thermal velocity is greater than drift velocity) and electron-driven; namely, the heating is applied directly to the electrons \cite{b:ahedo2010}. These electrons have strongly magnetised gyro-motion along the MN \cite{b:difede2021}; the cold ion trajectories are then determined by an ambipolar electric field that accelerates them downstream \cite{b:lafleur2014}. Thus, the MN radially confines the discharge, while converting electron thermal energy into directed axial ion kinetic energy, generating thrust \cite{b:martinez2015}. The two main realisations of the MEPT are the RF plasma thruster \cite{b:bellomo2021}---which includes the Helicon Plasma Thruster (HPT) \cite{b:takahashi2019helicon,b:lafleur2011,b:magarotto2019dep}---and the Electron Cyclotron Resonance Thruster (ECRT) \cite{b:correyero2019}.

\textcolor{black}{The absence of plasma-immersed electrodes or grids, as well as the magnetic confinement from the plasma source walls, means MEPTs are highly resistant to erosion \cite{b:magarotto2020_3}, which is often the lifetime-limiting aspect of conventional EP systems such as Ion and Hall effect thrusters (HET) \cite{b:goeb2008}. It is important to recognise though that advancements in magnetically-shielded (MS) HETs have effectively eliminated channel wall erosion \cite{mshall}. However, MS-HETs still suffer from cathode erosion, due to processes such as insert depletion and sputtering, as well as erosion at the magnetic poles \cite{Wang_2023,wang2}. While direct, long-duration experimental studies of MEPT lifetimes are still limited, the absence of significant erosion mechanisms provides a strong basis for their superior erosion resistance over even MS-HETs \cite{articleerosion}.}

MEPTs can also operate on a wider range of propellants, more advantageous to store for small spacecraft, and often inexpensive (e.g., iodine \cite{b:nabiliepc}, air \cite{b:andrewsabep} and water) \cite{b:bellomo2021}. MEPTs can also be less-complex than state-of-the-art devices, since they do not require a separate dedicated neutralising electron source (e.g. a hollow cathode), which consumes power and propellant \cite{b:lev}. Thus, MEPTs are becoming an increasing option for low-thrust propulsion, being highly scalable, robust, light, low-cost and resistant to erosion or complex failure \cite{b:manente2019}. However, most MEPT prototypes still report thrust efficiencies lower than 20\% \cite{b:takahashi2019helicon}.

While the main physical principles of MN's are relatively well understood, a complete description of the details of electron temperature evolution remains elusive. Measurements to date are evident that heat flux is significant in a MN expansion. Little and Choueri \cite{b:little2016} argued that adoption of classical heat conduction, i.e. a Fourier-like law governed by collisions, can explain experimental trends in electron cooling. However, it was noted that in practice, the magnitude of the classical collision frequency was too low, which led to a nonphysical heat flux downstream greater than the power delivered to the thruster. The authors thus concluded that even if the scaling of a Fourier law is appropriate, a much higher collision frequency would be required to physically explain the expansion.

This is of particular interest, since recent experimental work has shown that plasma instabilities may be a source for enhanced, non-classical collisionality in the MN plume \cite{b:hepner2020}. Given that low temperature ($T_e<20$~eV) MNs are typically characterised by strong cross-field gradients in potential and density, drift waves play a governing role \cite{b:hepner2020AIAA}. \textcolor{black}{Indeed, azimuthal plasma instabilities have been shown to lead to enhanced, "anomalous" cross-field transport in a variety of $E \times B$ discharges relevant to plasma propulsion \cite{b:exb}}, including HETs, and it was recently observed that a lower hybrid drift instability in a MN plume induces an effective collision frequency several orders of magnitude higher than the classical equivalent based on standard electron-neutral or electron-ion scattering \cite{b:hepner2020}. Such instabilities are driven by the large electron drift velocities in the azimuthal direction, which are often as fast as the electron thermal velocity itself, and which lead to large amplitude fluctuations in both the azimuthal electric field and the plasma density.

\textcolor{black}{Similar anomalous transport mechanisms have already been well-characterised in HETs. There exists a broad body of work that encompasses axial-azimuthal \cite{b:lafleuranom}, radial-azimuthal \cite{villafana}, and fully 3D simulations \cite{villafana3d,taccogna3d} of HET instabilities, as well as significant theoretical advancements in dispersion relations, instability criterion and induced transport \cite{hettheory1,hettheory2}. Compared to the extensive research on HETs, similar investigations into MNs are still in their infancy, leaving significant room for further exploration and development.}

Thus, given the role of instabilities in inducing an anomalous collision frequency higher than the classical value, and the inconsistency of classical collisions in MN expansion, the need is apparent to accurately model and analyse the applicability of an anomalous collision frequency in electron transport. Indeed, cross-field transport is critically linked to the process of detachment, whereby the expanded plasma ultimately decouples from the MN \cite{b:olsen2015}. Considering also that the electron heat flux directly influences the diamagnetic flow current, via the electron pressure gradient, accurate modelling of the anomalous transport is critical to giving correct numerical estimates of propulsive performance \cite{b:souhairAIP}.

Simple phenomenological \textcolor{black}{Bohm-type} models of anomalous transport have previously been applied within a magnetised electron fluid model, where the cross-field and parallel-field anomalous collisions were accounted for separately \cite{b:alvaro2021}. The simulated HPT exhibited an 18\% to 3\% drop in thrust efficiency when anomalous cross-field collisions were near the theoretical limit. \textcolor{black}{Sánchez-Villar et al. found that a hybrid model (with fluid electrons) had good agreement with experimental results of an ECR thruster when using certain values (0.035-0.08) of an empirical Bohm-like coefficient to include anomalous diffusion \cite{san}. Cichocki et al. applied the same best-fit approach to show good agreement between a simulation, with polytropic fluid electrons, and experimental measurements in the plume of a HPT for a coefficient of 0.02 \cite{b:cich}. Jiménez et al. used the same model as reference \cite{b:alvaro2021} in comparison to a cusped HPT \cite{jim}. Simulations were best-fit to a coefficient of 0.0165 and an anomalous parallel collisionality for heat flux of $O(10^7) Hz$.} However, the role of electron transport in the fluid treatment of the MN remains limited due to the absence of self-consistent closures for non-local heat conduction. \textcolor{black}{The ideal fluid model also neglects finite Larmor radius effects (gyro-viscous stress tensor and inertial terms), which have been shown to be comparable (circa 50\% of total local current density) to the combined magnitudes of $E\times B$ and diamagnetic terms in the fully kinetic Particle-in-Cell (PIC) study of Hu et al. \cite{b:Hu2021}. Indeed, in this article, it will be shown that gyro-viscous stress induced currents can be of similar value to $E\times B$ current in some conditions but certainly negligible in others.} Indeed, the PIC method represents the most self-consistent numerical strategy with the lowest level of assumptions. \textcolor{black}{The fully kinetic PIC method comes with considerably more computational burden than fluid or hybrid approaches, as well as inherent statistical noise. It typically requires scaling of physical parameters such as ion mass or vacuum permittivity to facilitate tractable computing time, which may conceal or affect the physics of interest. However, recent advances in these scaling methods, boundary conditions for small domains, and parallelisation have now enabled more efficient deployment of the PIC method \cite{b:andrews2022,b:levin2020,b:li2019}.}

A multiscale numerical model for MEPTs has recently been developed, coupling a 0D global model of the source region \cite{b:souhairPP,b:guaita2022} with a 2D fully kinetic PIC model for the MN \cite{b:andrews2022,b:lynden}. \textcolor{black}{In previous studies, best-fit values of a Bohm coefficient were found to range from 0.016 to 0.0625 for the 50 W-class and 150 W-class REGULUS thrusters, respectively \cite{b:andrewsiodine}.} While fully 3D PIC simulations can self-consistently capture plasma turbulence and azimuthal instabilities, 2D axisymmetric models do not resolve the azimuthal direction and therefore omit direct computation of the azimuthal electric field.

\textcolor{black}{This work aims to characterize the impact of anomalous effects on plasma transport and propulsive performance. Given this focus, a 2D PIC approach is sufficient, with anomalous diffusion incorporated through an equivalent electron scattering frequency rather than by explicitly resolving turbulence-driven instabilities. The goal is not to directly model the growth and saturation of conjectured instability modes, but rather to use the phenomenological Bohm model as a controlled scaling parameter. This enables a preliminary assessment of MN response to anomalous diffusion, providing a reference to the community that may inform future experiments or more advanced numerical studies.}

\textcolor{black}{This study represents the first fully kinetic analysis of anomalous diffusion effects on electron transport in an MN,} \textcolor{black}{using an effective electron collisionality based on the Bohm scaling model.} \textcolor{black}{Previous research has primarily focused on calibrating fluid-based simulations to experimental results \cite{b:alvaro2021,b:cich,san,jim}. Instead of employing anomalous diffusion coefficients purely for best-fit agreement with experiments, this work systematically explores the effect on plasma transport and the underlying mechanisms influencing MN performance.}

To this end, an outline of the coupled 0D global model and PIC methodology, as well as the theory of anomalous transport and the relation to azimuthal current and thrust is given in section 2. Section 3 then proceeds to provide a benchmark validation against independent experimental measurements. Section 4 provides parametric study of the MN of the 150 W class MEPT 'REGULUS-150' in terms of plasma profiles, azimuthal current formation, thrust generation and further propulsive performance metrics. Then, section 5 gives further insight into the role of anomalous collisionality in the electron cooling. Conclusions are given in section 6.

\section{Methodology}\label{sec:methodology}

\subsection{Particle-in-Cell model}
\begin{figure*}[!htb]
\includegraphics[width=\linewidth]{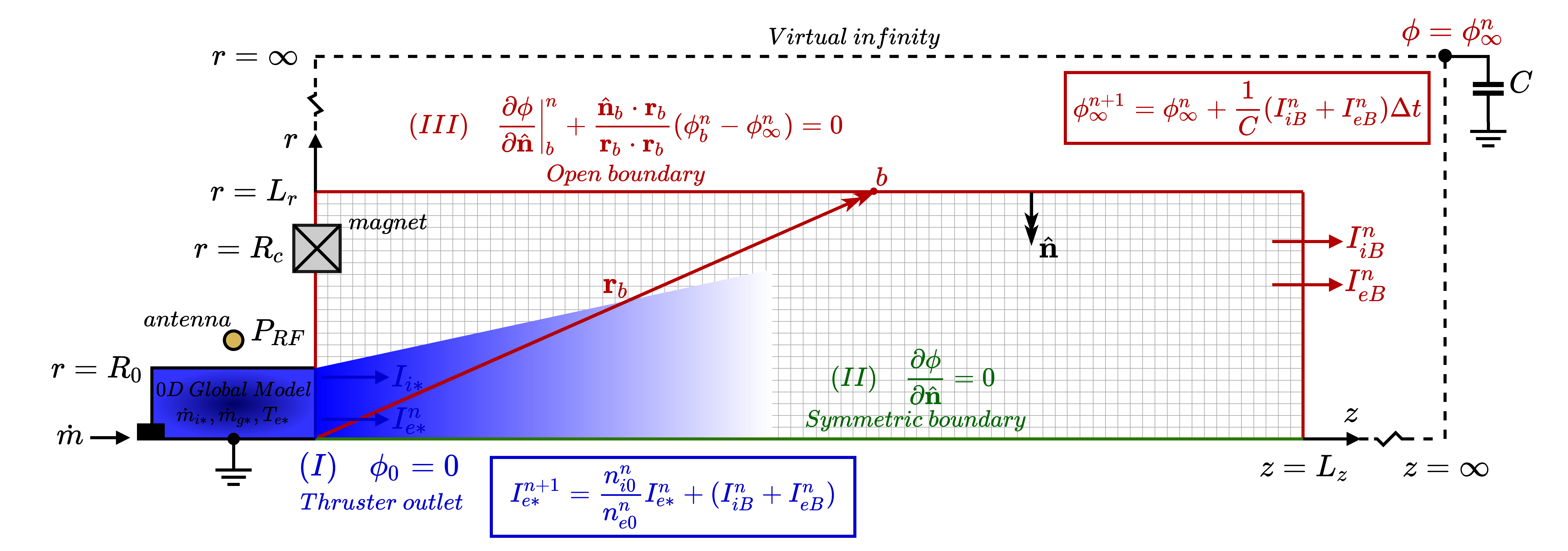}
\caption{\label{fig:domain}Simulation domain:\textcolor{black}{($I$) Thruster outlet, ($II$) symmetric boundary, and ($III$) open boundary. The magnet producing the MN is indicated at $r=R_c$. For the Poisson's equation, a Dirichlet condition applies to ($I$), zero-Neumann to ($II$), and Robin to ($III$). Ions and neutrals are absorbed on ($I$) and ($III$). Electrons are absorbed on ($I$) and selectively reflected on ($III$) based on an energy criterion. All particles are specularly reflected on ($II$). $I_{i*}$ and $I_{e*}$ are the injected ion and electron currents. $I_{iB}$ and $I_{eB}$ are the ion and electron currents lost to ($III$). $\phi_\infty$ is the free-space potential at infinity. $\mathbf{r}_b=\langle z_b,r_b,0\rangle$ is the outward vector from the centre of the thruster outlet to a point $b$ on the open boundary.}}
\end{figure*}
An axisymmetric 2D3V Particle-In-Cell Direct Simulation Monte Carlo (PIC-DSMC) model \cite{b:brieda2012} considers the axial $z$ and radial $r$ dimensions of partially-ionised plasma \textcolor{black}{(ions $i$, electrons $e$, neutrals $g$)} expansion in the MN. The simulation code is a development of that introduced extensively by Andrews et al. \cite{b:andrews2022}. In this section, only an overview of the model will be given, with particular detail on the additions made to include anomalous transport. The reader is therefore referred to reference \cite{b:andrews2022} for greater information.

The simulation domain is illustrated in \fref{fig:domain}. The plasma properties ($n_{e*}$, $n_{g*}$, $T_{e*}$) of the inductively coupled EM source discharge are evaluated using a 0D Global Model (GM), a complete description of which is provided in references \cite{b:souhairPP,b:guaita2022} and outlined in Appendix A. The GM is then coupled to the PIC model to provide the injection parameters at the thruster outlet (I), where ions, electrons and neutrals are injected. To ensure equal ion and electron current streams to the infinity at steady-state, the external boundaries (III) are treated as open to vacuum and stay connected to the thruster outlet via a virtual free-space capacitance. Boundary (II) is the axis of symmetry.

The subscripts $*$, $0$, $b$ and $\infty$ shall hereby refer to, respectively, properties in the plasma source (from the GM), at the thruster outlet boundary (I), at the open boundaries (III) and at the virtual infinity. The subscript B instead denotes the integral sum of local properties along the open boundaries.

\subsubsection{Particle-in-Cell}
Ions $i$, electrons $e$ and neutrals $g$ are all modelled as macro-particles, with their trajectory integration performed with the \textit{de facto} leap-frog Boris algorithm in equation \eref{eqn:boris}. Jim\'enez et al. recently showed that $>$95\% of deposited power is absorbed within the source tube, with $<$0.6\% absorbed in the far-plume   \cite{b:jimenez}; the EM power deposition in the MN is therefore assumed negligible and induced magnetic fields are not considered. 

An explicit successive over-relaxation (SOR) Gauss-Seidel scheme solves the self-consistent plasma potential $\phi$, according to the Poisson’s equation of equation \eref{eqn:poisson}:
\begin{equation}
\label{eqn:boris}
    \frac{\Delta\mathbf{v}_I^{n+1/2}}{\Delta t} = \frac{q_k}{f_km_k}\left( \mathbf{E}^n + \mathbf{v}^n_I \times \mathbf{B}\right),
    \qquad \frac{\Delta\mathbf{r}_I^{n+1}}{\Delta t} = \mathbf{v}_I^{n+1/2}\\
\end{equation}
\begin{equation}
\label{eqn:poisson}
    \gamma^2\varepsilon_0\nabla^2\phi=-\rho,
\end{equation}
\textcolor{black}{where $\mathbf{r}^n_I$ is the position of particle $I$ (belonging to the species subset $\mathcal{S}_k$, $k=i,e,g$), with mass $m_k$ and charge $q_k$ at time-step $n$ with velocity $\mathbf{v}^n_I$. $\Delta t$ is the time-step, $\mathbf{E}$ is the electric field, and $\mathbf{B}$ is the magneto-static field. $\rho=e(n_i-n_e)$ is the charge density, with $n_i$ the ion density, $n_e$ the electron density and $e=|q|$ the elementary charge. $\varepsilon_0$ is the vacuum permittivity of free-space.} 

\textcolor{black}{The constant factors $\gamma>0$ and $f_k<1$ scale the vacuum permittivity and heavy-species $k=i,g$ mass respectively ($f_e=1$) for the purpose of numerical acceleration \cite{b:szabo}.} Since temporal and spatial resolution of the plasma frequency $\omega_{pe}=\sqrt{n_ee^2/\varepsilon_0m_e}$ and Debye length $\lambda_D=\sqrt{\varepsilon_0k_BT_e/n_ee^2}$ is required, the permittivity scaling increases both mesh size and time-step by $\gamma$. The final result is a simulation speed-up of approximately $\gamma^2\sqrt{f}$. A more detailed description of the acceleration scheme and a sensitivity analysis on the values of $\gamma$ and $f$ used in the PIC model can be found in reference \cite{b:andrews2022}. At each time-step, the electric field is then updated per the electrostatic relation $\mathbf{E}=-\nabla\phi$.

\subsubsection{Boundary Conditions}

The thruster outlet (I) is given the reference potential $\phi_0=0$. A zero-Neumann condition ($\partial\phi/\partial\hat{\mathbf{n}}=0$) is applied to the symmetric $r=0$ boundary (II). At the open boundaries (III), the non-stationary Robin condition, established by Andrews et al. \cite{b:andrews2022}, is used:
\begin{equation}
    \frac{\partial\phi}{\partial\hat{\mathbf{n}}}\bigg|^n_b+\frac{\hat{\mathbf{n}}_b\cdot{\mathbf{r}_b}}{\mathbf{r}_b\cdot\mathbf{r}_b}\left(\phi^n_b-\phi^n_\infty\right) = 0,
    \label{eqn:robin}
\end{equation}
{where $\mathbf{r}_b$ is the vector from the centre of the thruster outlet (I) to a location on the open boundary (III), $\hat{\mathbf{n}}$ is the inward-pointing unit normal, and $\phi_\infty$ denotes the free-space plasma potential at infinity.}

At the thruster outlet (I), particles with a generalised Maxwellian distribution are injected each time step:
\begin{equation}
f_{I \in \mathcal{S}_k}^{+}(\mathbf{v}_{I}) = \sqrt{\frac{m_{k}}{2\pi k_B T_{k*}}}\exp \left(-\frac{m_{k}}{2k_B T_{k*}}|\mathbf{v}_{I}-\mathbf{u}_{k}|^2 \right)H(v_{Iz})
\label{eqn:f}
\end{equation}
{where $T_{k*}$ is the source reference species temperature. The drift velocity $\mathbf{u}_k$, imposed in the $z$ direction, is equal to the Bohm speed for ions and electrons: $\mathbf{u}_{i,e} = \langle c_{*},0,0\rangle$, where $c_{*}=\sqrt{k_B T_{e*}/m_i}$. \textcolor{black}{This is according to the requirement of a quasi-neutral sheath-like equilibrium at the source tube exit, where the plasma is choked at the MN throat. This has been confirmed by fluid simulations of the plasma production in the REGULUS thrusters \cite{b:magarotto2020_2,b:AA2022}. \textcolor{black}{Neutrals instead possess a drift velocity based on thermal effusion theory through an aperture $\mathbf{u}_{g} = \langle \bar{v}_g/4,0,0\rangle$, where $\bar{v}_g=\sqrt{8k_BT_{g*}/\pi m_g}$ for the Maxwellian}. $H()$ is the Heaviside function, since only forward-marching distributions ($v_{Iz}>0$) can be imposed.} \textcolor{black}{The radial density profiles of the injection are uniform, in accordance with what has been observed in simulations of the source tube of the REGULUS MEPTs \cite{b:magarotto2020_2,b:AA2022}. As such, it is expected for gradients to be concentrated near the edge.}

Ions, electrons and neutrals returning to the thruster outlet are absorbed, as are ions and neutrals reaching the open boundaries. For electrons at the open boundaries, an energy-based reflection criterion accounts for the trapped population returned by the ambipolar potential drop. The kinetic energy of electrons $KE_{e}$ is compared to the trapping potential $PE_b$ on the boundary node,
\begin{equation}
    KE_{e} = \frac{1}{2}m_e|\mathbf{v}_e|^2
\end{equation}
\begin{equation}
    PE_{b}=e(\phi_b-\phi_\infty).
\end{equation}

If $KE_{e}<PE_{b}$ the electron is trapped, so is reflected back with velocity $-\mathbf{v}_{e}$. Else, it is a free electron to be removed from the domain.

From this energy-based criterion, there is a value of $\phi_\infty$ that reflects the number of electrons required for a global current-free plume. Therefore, $\phi_\infty$ is self-consistently controlled via a virtual free-space capacitance $C$,
\begin{equation}
    \phi_\infty^{n+1}=\phi_\infty^n+\frac{1}{C}(I_{iB}^n+I_{eB}^n)\Delta t,
    \label{eqn:control}
\end{equation}
where $I_{iB}$ and $I_{eB}$ are the sum ion and electron currents leaving the open boundaries (III). The ion and electron currents streaming to infinity are therefore equal at steady-state.

To complete the circuit, any non-zero net current leaving the open boundaries ($III$) during the transient must be injected back into the domain through the thruster outlet ($I$). The injected electron current $I_{e*}$ is further controlled in order to enforce quasi-neutrality. While ions are injected with a constant current given by $I_{i*} = en_{i*}c_{*}A_0$, where $A_0$ is the area of the thruster outlet, the injected electron current is updated each time step according to 
\begin{equation}
{I_{e*}^{n+1}} = (I_{iB}^n+I_{eB}^n) + \frac{n_{i0}^n}{n_{e0}^n}{I_{e*}^n},
\label{eqn:QN}
\end{equation}
where the first term completes the circuit and the second term sustains quasi-neutrality. Since the injected electrons are Maxwellian, the initial current is ${I_{e*}^0}=-en_{*}(\bar{v}_{e*}/4+c_{*})A_0$. The injected neutral thermal flux is then $\Gamma_{g*}=n_{g*}\bar{v}_{g*}A_0/4$.

In conclusion, there exists a fully-consistent coupling between the current from the plasma source $I_{i0}=-I_{e0}$ to the open boundaries $I_{iB}=-I_{eB}$, the potential drop $\phi_\infty$ and the macroscopic plume solution. It is upheld via the boundary condition of equation~(\ref{eqn:robin}), the electron energy reflection condition, and the capacitive circuit control of equation~(\ref{eqn:control}) and equation~(\ref{eqn:QN}).

\subsubsection{Classical collisions}

\color{black}

The simulation takes into account Coulomb collisions, ion-neutral scattering, charge exchange collisions, and electron-neutral elastic and inelastic scattering. Intra-species collisions are treated using the complete Direct Simulation Monte Carlo (DSMC) technique \cite{b:bird1998}, while inter-species collisions are handled using the Monte Carlo Collision (MCC) methodology \cite{b:birdsall1991}. Regarding the ionisation interaction, only first ionisation has been considered. \textcolor{black}{The list of collisions implemented for xenon, and their respective forms of cross-section model, are provided in Table~\ref{Tab:Cross-sections}. The cross-sections are presented analytically, or as tabulations $T()$, as functions of $c_r$ and $E_r$, the relative collision speed and energy respectively of two particles $1,2$ with relative mass $m_r$. Coulomb collision cross sections are taken analytically, including self-consistent calculation of the Coulomb logarithm \cite{weng1990method}; the elastic scattering cross section is taken from the \texttt{Biagi v8.97} database \cite{biagi}; for ion-neutral and neutral-neutral elastic scattering, the analytical forms presented by Oh \cite{oh1997} and Szabo \cite{b:szabo} are used; The model of Rapp and Francis \cite{b:rapp1962} is used for ion-neutral charge exchange.}

\begin{table}[!htb]
\caption{\label{Tab:Cross-sections}\textcolor{black}{Interactions considered in the PIC model with associated sources for cross-section data.}}
\centering
\resizebox{\textwidth}{!}{
\begin{tabular}{cccc}
\hline
\textbf{Reactions} & \textbf{Reaction Type} & \textbf{Cross Section$^a$ [m$^2$]} & \textbf{Data Source}\\
\hline
$e + e \longrightarrow e + e$         & Coulomb scattering & $\sigma_{ee}=4\pi b_0^2\ln \Lambda$ & \cite{coulomb} \\
$e + Xe^+ \longrightarrow e + Xe^+$      & Coulomb scattering & $\sigma_{ei}=4\pi b_0^2\ln \Lambda$ & \cite{coulomb} \\
$e + Xe   \longrightarrow e + Xe$        & Elastic scattering & $\sigma_{en}=T(E_r)$ & \cite{biagi} \\
$e + Xe  \stackrel{{E^*=12.13 ~eV}}{\longrightarrow} 2e + Xe^+$       & Ionisation         & $\sigma_{iz}=T(E_r)$  & \cite{biagi,hyman1979} \\
$e + Xe  \stackrel{{E^*=8.50 ~eV}}{\longrightarrow} e + Xe^*_{1S_M}$  & 1s$_{5,3}$ Metastable Excitation & $\sigma_{ex,1S_M}=T(E_r)$ &  \cite{b:souhairPP,biagi,gangwar2019}\\
$e + Xe  \stackrel{{E^*=9.00 ~eV}}{\longrightarrow} e + Xe^*_{1S_R}$  & 1s$_{4,2}$ Resonant Excitation & $\sigma_{ex,1S_R}=T(E_r)$ &  \cite{b:souhairPP,biagi,gangwar2019}\\
$e + Xe   \stackrel{{E^*=10.17 ~eV}}{\longrightarrow} e + Xe^*_{2P}$   & 2p$_{10-1}$ Excitation &  $\sigma_{ex,2P}=T(E_r)$&  \cite{b:souhairPP,biagi,zatsarinny2004,gangwar2019}\\
$Xe^+ + Xe \longrightarrow Xe^+ +Xe$      & Elastic Scattering & $\sigma_{{in,el}} = {8.28072 \times 10^{-16}}{c_r^{-1}}$ & \cite{oh1997,b:szabo} \\ 
$Xe^+ + Xe \longrightarrow Xe + Xe^+$     & Charge exchange    & $\sigma_{in,cex}=(15.1262-0.8821\ln c_r)^2\times 10^{-20}$ & \cite{rapp1962} \\
$Xe + Xe   \longrightarrow Xe + Xe$       & Elastic scattering & $\sigma_{nn}=2.117\times 10^{-18}c_r^{-0.24}$ & \cite{oh1997,b:szabo} \\
\hline
\end{tabular}}
\raggedright $^a$ $b_0=\frac{e^2}{4\pi\varepsilon_0 m_r c_r^2}, \quad m_r=\frac{m_1m_2}{m_1+m_2}, \quad c_r=v_2-v_1, \quad E_r=\frac{1}{2}m_rc_r^2, \quad \Lambda = \left( 1+\left(\frac{\lambda_D}{b_0}\right)^2 \right)^{\frac{1}{2}}$
\end{table}

\begin{figure}[!htb]
\centering
\includegraphics[width=\linewidth]{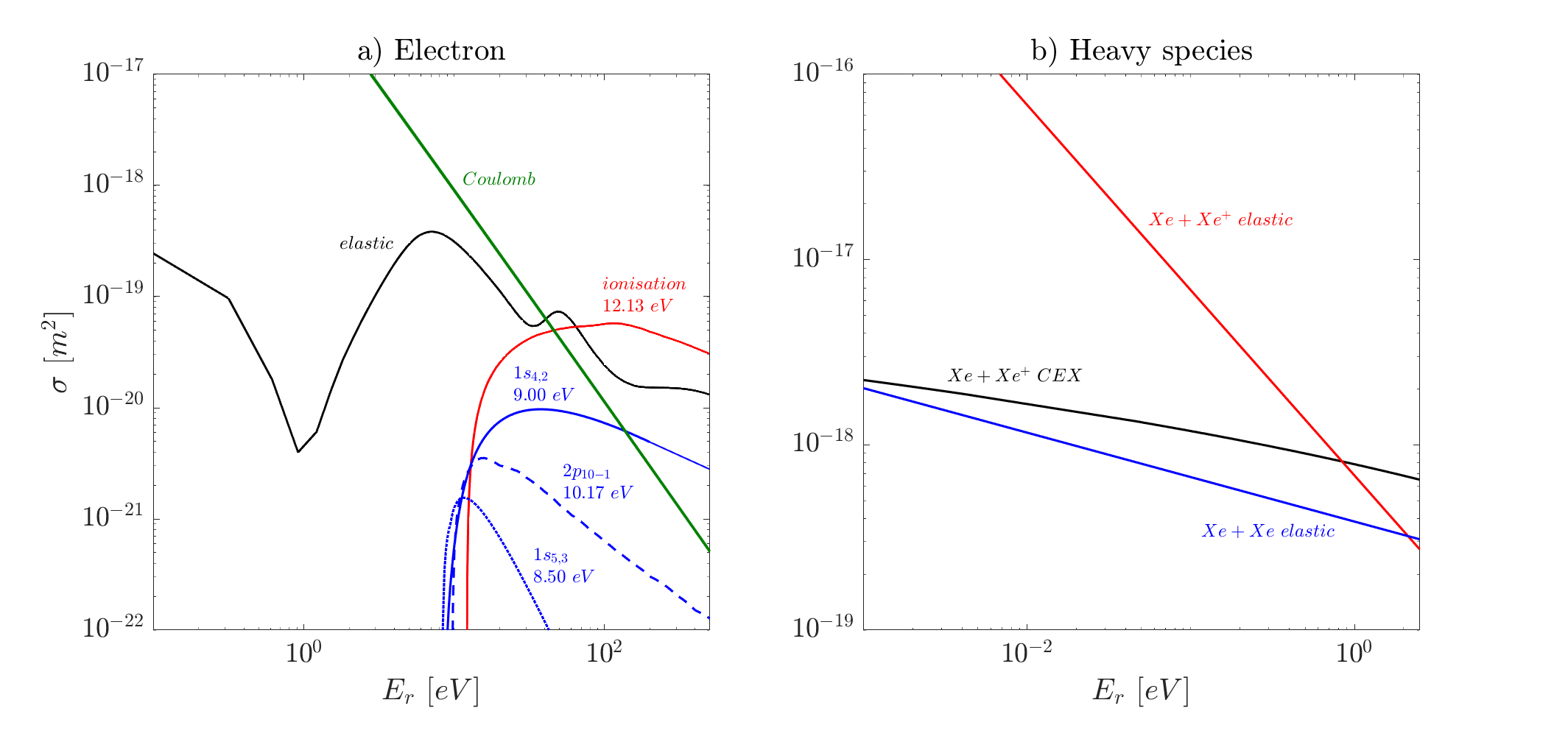}
\caption{\label{fig:crosssection} \textcolor{black}{(a) Electron collision cross-sections, including the lumped states for excitation (Coulomb cross-section calculated at $n_e=10^{18}$ m$^{-3}$); (b) Heavy species collision cross-sections.}}
\end{figure}

\textcolor{black}{Considering a full fine-structure collisional-radiative model of xenon's collisional excitation and ionisation reactions, and radiative decay of those excited states, would involve the consideration of 64 electron-neutral MCC processes per time-step (considering only the mechanisms between the $1s$ and $2p$ excited states, as the typical operating pressure of a MEPT falls within the range where these levels are predominantly excited). Treatment would then also be required for 14 excited species besides the ion and ground states. This leads to an unmanageable computational requirement. Thus, in order to reduce the number of collision probabilities to evaluate, a lumping procedure based on the assumption of local thermodynamic equilibrium is used, introduced in the previous work of Souhair et al. \cite{b:souhairPP}. Therefore, the 14 excited levels of the fine structure have been grouped into three lumped levels: regarding the $1s$ state, both resonant (i.e., $1s_4$ and $1s_2$) and metastable (i.e., $1s_5$ and $1s_3$) energy levels are lumped into $1S_R$ and $1S_M$. All ten levels of the $2p_{10-1}$ group are then lumped into a single state $2P$. Ionisation from excited states is also lumped into the ionisation from ground state. The procedure defined in reference \cite{b:souhairPP} for reaction rates is adapted here to give the lumped collision cross section as,}

\begin{equation}
    \sigma_{IJ} = \sum^{N_i}_i\left[\frac{\sum^{N_j}_j\sigma_{ij}}{\sum_{m}^{N_i}\frac{g_m}{g_i}exp\left(-\frac{\Delta E_{im}}{k_BT_g}\right)}\right]
\end{equation}

for the lumped state $I$ to $J$, where $i$ and $j$ span the fine structure of the lumped states in that group of $N_i$ and $N_j$ levels and $m$ is an index of the generic levels of the fine-structure. $\sigma_{IJ}$ is the lumped cross section, and $\sigma_{ij}$ is original cross section of the fine-structure reaction. $\Delta E_{im}=E_m-E_i$ is the energy gap between the two states, and $g_i$ and $g_m$ are the statistical weights that represent the degeneracy of, respectively, the lower and the higher energy level. $T_g$ is the neutral gas temperature. The lumped excitation energies are then given as,

\begin{equation}
    E^*_I = \frac{\sum_i^{N_i}g_iE_i}{\sum^{N_i}_ig_i}.
\end{equation}

Post-collision, the excited species are not modelled. Instead, it is assumed that excited neutrals undergo immediate radiative decay to the ground state. Therefore, the excitation collisions effectively represent a pure electron energy sink. The ionisation from ground state cross section is taken from the \texttt{Biagi v8.97} database \cite{biagi}, while those from the excited states are taken from Hyman \cite{hyman1979}. The excitation cross sections from the ground state are found from both the \texttt{Biagi v8.97} \cite{biagi} and the BSR database that uses the quantum calculations of Zatsarinny and Bartschat \cite{zatsarinny2004}. Step-wise excitation ($1s \rightarrow 1s$ and $1s \rightarrow 2p$) cross sections of the fine structure are taken from Priti et al. \cite{gangwar2019}. The complete set of the final cross-section data used in the DSMC/MCC module of the simulation is given, in terms of $E_r$, in \fref{fig:crosssection} (a) and (b) for the electron and heavy species reactions respectively. 

\color{black}

\subsubsection{Anomalous collisions}
The model of the anomalous collision frequency is crucial for the resulting plasma dynamics. In the previous decade, many attempts have been made to find accurate, flexible and self-consistent models. The empirical Bohm model \cite{b:Bohm} is based on the idea that anomalous Bohm diffusion results in the additional electron mobility observed experimentally. \textcolor{black}{\st{The theory behind the mechanism for Bohm diffusion, presented by Esipchuck et al., considers that the azimuthal drift waves can exist only in regions with decreasing gradients in the magnetic field} Bohm suggested that the rate of diffusion was linear with temperature and inversely linear with the strength of the confining magnetic field}. The anomalous collision frequency is
\begin{equation}
    {\nu}_{an} = \alpha_{an}\omega_{ce}
\end{equation}
where $\alpha_{an}$ is an empirical coefficient and $\omega_{ce}=e|\mathbf{B}|/m_e$ is the local electron gyro-frequency. In the fully-turbulent limit, $\alpha_{an}=1/16$, \textcolor{black}{though previous studies of MNs in MEPTs have found best fit values between 0.01-0.1 \cite{b:alvaro2021,b:cich,san,jim}. It should be emphasised that recent experimental and numerical investigations of HETs does not support the application of constant Bohm collisionality models \cite{b:goeb2008}. Already in the last decade, multi-region Bohm models were employed and only very-recently has work commenced on self-consistent anomalous collision models derived from first-principle instability theory \cite{b:lafleuranom,hettheory1}. The Bohm model here is applied as a suitable way to achieve preliminary parametric study into the MN under anomalous diffusion}. The MCC probability of an anomalous collision event is then
\begin{equation}
    P_{an} = 1-exp(-\nu_{an}\Delta t)
\end{equation}
If a collision occurs, electrons are assumed to scatter preferably in the direction perpendicular to the magnetic field \textcolor{black}{\cite{b:Bohm}}. Therefore, the post-collision velocity $\mathbf{v}_e^{\prime}$ is derived by rotating the perpendicular component of the electron velocity vector $\mathbf{v}_{e\perp}$ with respect to $\mathbf{B}$ with an arbitrary angle $\psi=2\pi\mathcal{R}[0,1]$, where $\mathcal{R}[a,b]$ is a random number between $a$ and $b$,and leaving the parallel component unchanged.
\begin{equation}
\mathbf{v}_e^{\prime}=\underbrace{\left( \mathbf{v}_e\cdot \hat{\mathbf{b}} \right)\hat{\mathbf{b}}}_{\mathbf{v}_{e\parallel}} + \mathbf{Q}_\perp(\psi)\underbrace{\left( \mathbf{v}_e -  \left( \mathbf{v}_e\cdot \hat{\mathbf{b}} \right)\hat{\mathbf{b}} \right)}_{\mathbf{v}_{e\perp}}
\end{equation}
where
\begin{equation}
   \mathbf{Q}_\perp(\psi) =  \left[ \begin{array}{ccc}
cos(\psi) & 0 & sin(\psi)\cr
0 & 1 & 0 \cr
-sin(\psi) & 0 & cos(\psi)
\end{array}\right]
\end{equation}
is the perpendicular rotation matrix \textcolor{black}{in the vector basis of $\hat{\mathbf{b}}$, where} $\hat{\mathbf{b}}=\mathbf{B}/|\mathbf{B}|$.

\subsection{Electron transport}\label{sec:collisional}
The parameters discussed in the following are calculated in post-processing as moments of the distribution functions provided by the PIC model \cite{b:chen1984}. \textcolor{black}{For instance, the 0th and 1st moments give the number density and drift velocity respectively as,}

\begin{equation}
\textcolor{black}{n_k \equiv \int f_k(\mathbf{v_k})~d\mathbf{v}_k,}
\end{equation}

\begin{equation}
\textcolor{black}{\mathbf{u}_k \equiv \frac{1}{n_k}\int \mathbf{v}_k f_k(\mathbf{v_k})~d\mathbf{v}_k.}
\end{equation}

\textcolor{black}{By taking the fluid moments of the kinetic equations, it is possible to express relevant properties as a sum of contributing terms. This allows greater physical insight rather than the limitations of total properties calculated from particles. It is highlighted that no fluid assumptions are made and that all quantities are derived directly from the PIC distributions.}

\subsubsection{Momentum, azimuthal current and thrust}
At steady-state, the first integral moment of the electron velocity distribution function fulfills the momentum equation \cite{b:bittencourt2004}, 
\begin{eqnarray}
    \mathbf{0} = en_e(\mathbf{E}+\mathbf{u}_e\times\mathbf{B}) +\mathbf{\nabla} \cdot (p_e\rttensor{\mathbf{{I}}} + \rttensor{\bm{\pi}}_e + m_en_e\mathbf{u}_e\mathbf{u}_e) - \mathbf{R}_{e} - \mathbf{R}_{an}, 
\end{eqnarray}
where: $\mathbf{u}_e$ is the electron macroscopic drift velocity; the (symmetric) pressure tensor \textcolor{black}{is given by the 2nd moment of the PIC kinetic distribution,}
\begin{equation}
 \rttensor{\mathbf{P}}_e \equiv m_e\int\mathbf{w}_e\mathbf{w}_ef_e(\mathbf{v_e})~d\mathbf{v}_e =p_e\rttensor{\mathbf{I}}+\rttensor{\bm{\pi}}_e,
\end{equation}
with $\rttensor{\mathbf{I}}$ the identity tensor and $\mathbf{w}_e=\mathbf{v}_e-\mathbf{u}_e$, has been split into the scalar static isotropic part $p_e=\frac{1}{3}tr(\rttensor{\mathbf{P}}_e)= k_Bn_eT_e$ and the anisotropic viscosity tensor $\rttensor{\bm{\pi}}$; $\mathbf{R}_{e}\equiv\mathbf{R}_{ei}+\mathbf{R}_{en}$ is the resistive collisional friction force density due to electron collisions with species $k =i,n$
\begin{equation}
     \mathbf{R}_{e} = -\sum_{k}m_{ek}n_k(\mathbf{u}_e-\mathbf{u}_k)\nu_{ek}
\end{equation}
with $\nu_{ek}$ the collision frequency and $m_{ek}=m_em_k/(m_e+m_k)$ the reduced mass; the anomalous term $\mathbf{R}_{an}$ can be considered an equivalent instability-enhanced electron-ion friction force density,
\begin{equation}
    \mathbf{R}_{an} = -m_en_e(\mathbf{u}_e-\mathbf{u}_i)\rttensor{\bm{\nu}}_{an}
\end{equation}
where $\rttensor{\bm{\nu}}_{an}$ is the diagonal anomalous collision frequency tensor---written in this form only to imply that it is not necessarily isotropic---and where $\mathbf{1}_\theta\cdot\rttensor{\bm{\nu}}_{an}\cdot\mathbf{1}_\theta=\nu_{an}$. \textcolor{black}{The anomalous collisions are also expressed in tensor form to retain the generality that anomalous diffusion and/or heat transfer may well be naturally occurring in the axial-radial plane of the simulation (for example modified two-stream instabilities parallel to $\mathbf{B}$).} The anomalous axial and radial collision frequency is then instead obtained from the conductivity tensor $\rttensor{\bm{\sigma}}=\frac{n_ee^2}{m_e}\cdot(\rttensor{\bm{\nu}}_{an}^{\odot -1}+\nu_e^{-1})$ \cite{b:bittencourt2004} \textcolor{black}{($\odot$ is the Hadamard element-wise operation)}, which is obtained from Ohm's law $\mathbf{j}_e=\rttensor{\bm{\sigma}}(\mathbf{E}+\mathbf{u}_e\times\mathbf{B})$ \textcolor{black}{and where the classical $\nu_e$ is kinetically obtained from the MCC/DSMC module of the simulation.} \textcolor{black}{Note that the rotation of the perpendicular velocity components (both azimuthal and normal) in the anomalous collision model is a necessary consequence of ensuring energy and momentum conservation for what are taken as purely azimuthal collisions. The modifications to the perpendicular velocity do not introduce independent scattering effects but are instead required to maintain only a consistent gyro-velocity update.}

\textcolor{black}{In a mesothermal plasma (i.e. $\mathbf{u}_e << \mathbf{v}_{th,e}$) the inertial component of the substantive derivative can be typically considered negligible ($\mathbf{\nabla}\cdot(m_en_e\mathbf{u}_e\mathbf{u}_e)\sim 0$), and this will be shown in Section 4. However, it will be retained for now in the derivation for completeness}. The total flow components are then obtained by taking the cross product of the momentum equation with the magnetic field $\mathbf{B}$:
\begin{eqnarray}
\label{eqn:flowvel}
     \mathbf{u}_e &= \frac{(\mathbf{B}\cdot\mathbf{u}_e)\mathbf{B}}{|\mathbf{B}|^2} + \frac{\mathbf{E}\times\mathbf{B}}{|\mathbf{B}|^2} + \frac{\mathbf{\nabla}p_e\times\mathbf{B}}{en_e|\mathbf{B}|^2} + \frac{\mathbf{\nabla}\cdot\bm{\pi}_e\times\mathbf{B}}{en_e|\mathbf{B}|^2} \nonumber \\
     &\qquad + \frac{\mathbf{R}_{e}\times\mathbf{B}}{en_e|\mathbf{B}|^2} + \frac{\mathbf{R}_{an}\times\mathbf{B}}{en_e|\mathbf{B}|^2} + \textcolor{black}{\frac{\mathbf{u}_e\nabla\cdot\mathbf{u}_e\times \mathbf{B}}{|\mathbf{B}|^2}}.
\end{eqnarray}
The first term in equation \eref{eqn:flowvel} is the parallel flow velocity $\mathbf{u}_{\parallel}=\mathbf{u}_e \cdot \mathbf{B}/|\mathbf{B}|$. Making use of the fact that $\mathbf{u}_\parallel \cdot\mathbf{1}_\theta = 0$, the azimuthal electron current density $j_{e\theta}$ is then
\begin{eqnarray}
    j_{e\theta} = -en_eu_{e\theta} &= -\left[
\underbrace{en_e\frac{\mathbf{E}\times\mathbf{B}}{|\mathbf{B}|^2}}_{{E}\times{B}} 
+ \underbrace{\frac{\mathbf{\nabla}p_e\times\mathbf{B}}{|\mathbf{B}|^2}}_{\mathrm{Diamagnetic}} 
+ \underbrace{\frac{\mathbf{\nabla}\cdot\bm{\pi}_e\times\mathbf{B}}{|\mathbf{B}|^2}}_{\mathrm{Stress}}\right. \nonumber \\
&\qquad + \left. \textcolor{black}{\underbrace{\frac{(\mathbf{R}_{e}+\mathbf{R}_{an})\times\mathbf{B}}{|\mathbf{B}|^2}}_{\mathrm{Friction}}}+\textcolor{black}{\underbrace{\frac{\mathbf{u}_e\nabla\cdot\mathbf{u}_e \times \mathbf{B}}{|\mathbf{B}|^2}}_{\mathrm{Inertial}}} 
 \right] \cdot \mathbf{1}_\theta.
     \label{eqn:azimuthalcurrent}
\end{eqnarray}
In equation \eref{eqn:azimuthalcurrent}, the terms on the right-hand side correspond to the contributions due to the $\mathbf{E}\times\mathbf{B}$ drift $j_{e\theta}^{E\times B}$, diamagnetic flow $j_{e\theta}^{\chi}$, gyro-viscous stress $j_{e\theta}^{\pi}$, friction $j_{e\theta}^{f}$, \textcolor{black}{and inertial term $j_{e\theta}^{M}$ respectively}.

The total axial momentum flux $F$ (thrust) of the plasma expansion is 
\begin{equation}
    F = \oiint\limits_{S(V)} \sum_{k=i,e,g} \mathbf{1}_z\cdot\left(m_kn_k \mathbf{u}_k\mathbf{u}_k + p_k\mathbf{\rttensor{I}} +\bm{\rttensor{\pi}}_k\right)\cdot \hat{\mathbf{n}} ~dS
    \label{eqn:force}
\end{equation}
where $S$ is the bounding surface of the simulation control volume $V$. Considering the summation of the electron momentum equation with its ion and neutral counterparts, the thrust can also be expressed by
\begin{equation}
    F = F_0 + \iiint\limits_V -j_{e\theta}B_r~dV
\end{equation}
where $F_0$ is the axial force entering the MN from the plasma source \cite{b:lafleur2015}. The integrand term can be considered as the Lorentz force density, and can be separated into the same contributions as described for $j_{e\theta}$ in equation \eref{eqn:azimuthalcurrent}. \textcolor{black}{Note that ion azimuthal current is negligible \cite{b:lafleur2014,b:merino2016}; Ions have negligible azimuthal velocity because their large mass results in a very-low gyro-frequency. They can be treated as unmagnetised and significant azimuthal motion cannot occur in the absence of a strong azimuthal electric field or high initial perpendicular velocity (ions are injected cold)}. This can be then decomposed into the azimuthal current contributions listed in equation \eref{eqn:azimuthalcurrent},

\begin{equation}
     F_{E\times B} = \iiint\limits_V -j_{e\theta}^{E \times B}B_r~dV,
\end{equation}
\begin{equation}
     F_{\chi} = \iiint\limits_V -j_{e\theta}^{\chi}B_r~dV,
\end{equation}
\begin{equation}
     F_{\pi} = \iiint\limits_V -j_{e\theta}^{\pi}B_r~dV,
\end{equation}
\begin{equation}
     F_{f} = \iiint\limits_V -j_{e\theta}^{f}B_r~dV,
\end{equation}
\begin{equation}
     \textcolor{black}{F_{M} = \iiint\limits_V -j_{e\theta}^{M}B_r~dV.}
\end{equation}

\subsubsection{Power and efficiency}

\textcolor{black}{The steady-state energy equation for each species $k$ is obtained by taking the energy moment of the kinetic equation \cite{b:bittencourt2004}:}
\begin{eqnarray}
    \mathbf{0} = & \mathbf{\nabla}\cdot\left[\mathbf{q}_k + \left( \frac{5}{2}k_Bn_kT_k+\frac{1}{2}n_km_k|\mathbf{u}_k|^2 \right)\mathbf{u}_k + \mathbf{u}_k\cdot\rttensor{\bm{\pi}} \right] \nonumber \\ &  - n_kq_k\mathbf{u}_k\cdot\mathbf{E} + \sum_{s\neq k}\left(U_{ks} -\mathbf{u}_k\cdot\mathbf{R}_{ks} - \Psi_{ks}\right)
\end{eqnarray}
The divergence term includes the sum of the net energy flux densities into the local volume due to heat conduction $\mathbf{q}$, \textcolor{black}{given by the 3rd moment of the kinetic PIC distribution,}

\begin{equation}
   \textcolor{black}{\mathbf{q}_k = \frac{m_k}{2} \int \mathbf{w}_k |\mathbf{w}_k|^2 f_k(\mathbf{v}_k) d\mathbf{v}_k,}
\end{equation}

\textcolor{black}{heat convection, macroscopic kinetic, and dissipation due to flow-gradient-induced stress respectively. The second term is the electric field energy.  The remaining terms account for the work done by collisional heat-exchange $U_{ks}=-n_k\frac{m_k}{m_s}\nu_{ks}\frac{3k_B}{2e}\left(T_s-T_k\right)$, frictional heat-exchange, and inelastic collision energy losses such as ionisation and excitation $\Psi$, summed over species $s\neq k$ \cite{b:goeb2008}.}\\
Taking the volume integral of equation 21, the terms of the MN power balance $P_*=\sum_k\left(P_{kin,k}+P_{P,k}+Q_k\right)+P_{loss}$ can be defined, \textcolor{black}{where again $S$ is the bounding open surface of the simulation control volume $V$}, as:
\begin{eqnarray}
    P_{kin,k} &= \int_S \left[\frac{1}{2}n_km_i|\mathbf{u}_k|^2\mathbf{u}_k \right]\cdot\hat{\bm{n}}\ dS, \\ 
    P_{P,k} &= \int_S \left[\frac{5}{2}n_kT_k\mathbf{u}_k + \mathbf{u}_k\cdot\rttensor{\bm{\pi}}_k\right] \cdot\hat{\bm{n}}\ dS, \\
    Q_{k}   &= \int_S \mathbf{q}_k \cdot\hat{\bm{n}}\ dS, \\
    P_{loss} &= \int_V \sum_{k}\left(U_{k}+\mathbf{u}_k\cdot\mathbf{R}_{k}+\Psi_{k}\right)~dV, \\
    P_* &= \sum_k\left[ P_{kin,k}(0)+P_{P,k}(0)+Q_k(0) \right].
\end{eqnarray}
These are the macroscopic kinetic energy flux, total pressure energy flux, heat flux, collisional loss power and the total power entering the MN respectively. \textcolor{black}{Note that in $P_{loss}$, the sum of elastic heat transfer across species cancels out, but each term is retained for completeness, since these terms can be included as losses when a separate background facility gas is considered \cite{b:andri}.}. Power-related efficiencies of interest are the energy conversion efficiency, the divergence efficiency and the total MN efficiency, which can then be defined as:

\begin{eqnarray}
    \color{black} \eta_{ene} &=& \color{black} \frac{\sum_k \left (P_{kin,k}+P_{P,k} \right)}{P_*},  \\  
    \color{black} \eta_{div} &=& \color{black} \frac{\sum_k \left (P^{(z)}_{kin,k}+P^{(z)}_{P,k} \right)}{\sum_k \left(P_{kin,k}+P_{P,k} \right)},\\  
    \color{black} \eta_{MN} &=& \color{black} \frac{\sum_k \left (P^{(z)}_{kin,k}+P^{(z)}_{P,k} \right)}{P_*} = \eta_{ene}\eta_{div}.
\end{eqnarray}
\textcolor{black}{The superscript $(z)$ refers to the axial component of the power term. These efficiencies represent the effectiveness of electron thermal to ion kinetic energy conversion, plume confinement, and overall MN thrust generation \cite{b:ahedo2010,jim}.} The typical thrust efficiency can be given as
\begin{equation}
     \eta_F = \frac{F^2}{2\dot{m}P_{a}}\approx \eta_s\eta_{MN}, 
\end{equation}
where $\eta_s=P_*/P_a$ is the source efficiency provided by the global model (see \ref{sec:appendixA}).

\section{Independent benchmark validation}

\begin{table}[h]
    \centering
    \renewcommand{\arraystretch}{1.2}
    \caption{\textcolor{black}{Simulation parameters: Independent benchmark validation}}
    \begin{tabular}{llccc}
        \toprule
        \textbf{Parameter} & \textbf{Symbol} & \textbf{Unscaled} & \textbf{Scaled} & \textbf{Units} \\
        \midrule
        Outlet Radius & $R_0$ & 37.5 & - & [mm] \\
        Ion Mass$^a$ ($Ar$) & $m_i$ & 40 & 1 & AMU \\
        Reference Plasma Density & $n_{i*,e*}$ & $10^{18}$ & - & m$^{-3}$ \\
        Reference Neutral Gas Density & $n_{g*}$ & $10^{17}$ & - & m$^{-3}$ \\
        Reference Electron Temperature & $T_{e*}$ & 7 & - & eV \\
        Reference Ion/Neutral Temperature & $T_{i*,g*}$ & 298 & - & K \\
        Ion Injection Speed$^a$ & $u_{iz0}$ & 1.12$c_*$ & 7.1$c_*$ & - \\
        Magnet Radius & $R_c$ & 75.1 & - & [mm] \\
        Throat Magnetic Field Strength & $B_0$ & 370 & - & [G]\\
        \midrule
        Axial Domain Length & $L_z$ & $13R_0$ & - & - \\
        Radial Domain Length & $L_r$ & $6R_0$ & - & - \\
        Reference Debye Length$^a$ & ${\lambda_{D*}}$ & 0.0197 & 1.875 & mm \\
        Time-Step & $\Delta t$ & $1 \times 10^{-10}$ & - & s \\
        Capacitance & $C$ & 0.4 & - & nF \\
        \bottomrule
    \end{tabular}\\
    \raggedright $^a$ $f=40$, $\gamma=95$.
    \label{tab:params}
\end{table}

\begin{figure}[!htb]
\centering
\includegraphics[width=0.6\linewidth]{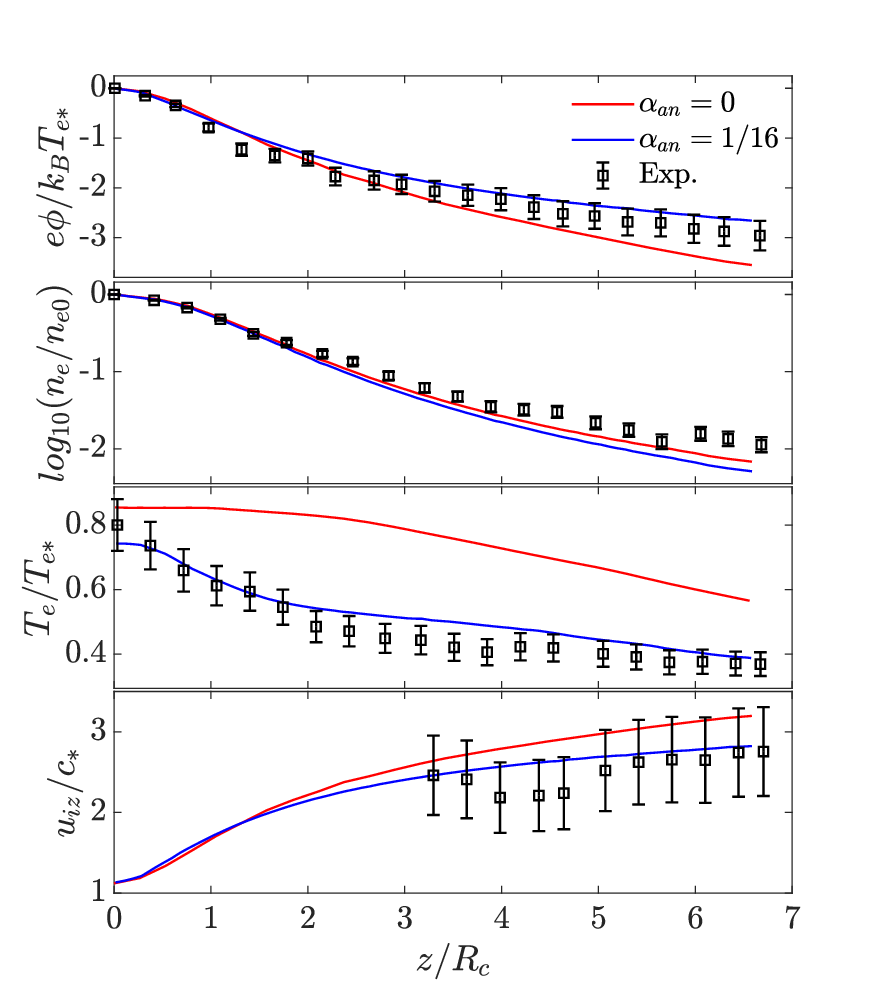}
\caption{\label{fig:valid} Benchmark validation of the PIC model, with anomalous transport, against the experimental data measured by Little and Choueiri \cite{b:little2016}.}
\end{figure}
Although the PIC model has been validated previously \cite{b:andrews2022} with the measurements of Lafleur et al. \cite{b:lafleur2011}, the magnetic field was too weak ($B_0\approx 40$~G) for the anomalous transport to be significant compared to classical collisions. The experimental results from the RF thruster of Little and Choueiri \cite{b:little2016} have become a \textit{de facto} benchmark for MN simulations \cite{b:chen2020,b:chen2022}.
The \textcolor{black}{centreline} measurements shown in \fref{fig:valid} were gathered for an Argon mass flow rate of $\dot{m} = 0.5$~mg/s, \textcolor{black}{into a source tube of radius $R_0=$ 37.5 mm and length 305 mm, with the power delivered to the antenna of approximately 500~W; the MN throat strength, provided by a pair of concentric magnets with radius $R_c=$ 75.1 mm, was $B_0\approx 370$~G. This resulted in reference properties of $n_{e*}\approx 10^{18}$~m$^{-3}$ and $T_{e*} \approx 7$~eV, with $n_{g*}\approx 10^{17}$~m$^{-3}$ estimated from the reported 20 $\mu$Torr (2.67$\times$10$^{-3}$ Pa) pressure}. A swept RF-compensated Langmuir probe (LP) measured the plasma density and electron temperature; the plasma potential was obtained with a heated emissive probe (EP) using the floating point method; the ion velocity and ion energy distribution function were measured with a four-grid retarding potential analyzer (RPA). The uncertainty bands associated to the LP are 5\% and 10\% for density and temperature respectively. The error of the EP is 5\%, and the RPA 25\%.

\textcolor{black}{The domain was taken to be $L_z=13R_0$, $L_r=6R_0$, with uniform mesh spacing of $\Delta z = \Delta r=$1.875 mm. This was facilitated with a permittivity scaling of $\gamma=95$, such to resolve the Debye length $\lambda_D$. Argon ions and neutrals were given a mass scaling of $f=40$. The time-step, constrained by the electron Courant-Friedrich-Lewy condition, was $\Delta t=1\times10^{-10}$ s. The macro-particle weigh was chosen such that the minimum number of charged particles per cell (per species) was $(N_p)_{min}>16$, and $C=0.4$ nF, both per the sensitivity studies in reference \cite{b:andrews2022}. Extended discussion regarding the choice of numerical parameters is offered in section 4.1.3 concerning the setup of the main body of simulations in this work. Ion and neutrals were injected cold, that is $T_{i*}=T_{g*}=$ 298 K. All species were injected with uniform radial density profiles. Since the experiment featured a MN throat located behind the physical source tube exit, the ions are injected with the actual measured drift velocity at the exit of $u_{iz0}=1.12c_*$. A summary of the relevant numerical parameters to the validation simulation are given in table \ref{tab:params}}. \textcolor{black}{In this high-power case, the mass utilisation is high and, with the neutral gas density low $<10^{17}$ m$^{-3}$, the Argon heavy-species mean free paths were estimated to be $\lambda \approx 10^0 - 10^2$ m and thus collisions involving neutrals can be neglected. Coulomb collisions are included as per the model described in Section 2.1.3}.

The comparisons of the simulated and measured properties along the MN axis are presented in \fref{fig:valid}, where both classical and fully-turbulent anomalous transport ($\alpha_{an}=1/16$) have been considered, \textcolor{black}{the latter found to be the best-fit coefficient from a range of tested $\alpha_{an}$}. There is excellent agreement in the fully-turbulent case, particularly in the plasma potential of \fref{fig:valid} (a), ion axial velocity \fref{fig:valid} (d), and electron temperature \fref{fig:valid} (c), where the classical transport model fails to capture the rapid cooling seen experimentally. \textcolor{black}{Anomalous diffusion leads to a much higher electron cross-field mobility compared to classical (collisional) transport. Since electrons---particularly those of high-energy----move more freely across $\mathbf{B}$, they more effectively dissipate their energy, reducing the electron cooling length scale as the plasma expands. Regarding plasma potential, the increased cross-field transport results in greater plume divergence and earlier electron detachment; the effective escape energy of the free electron population is reduced, and a weaker potential drop is thus required to maintain the current-free condition. Accordingly, there is slower ion acceleration.} It must be acknowledged that the electron density profile of \fref{fig:valid} (b) has increased discrepancy in the fully-turbulent case, but the error is still $<15$\%. The reasons for this are most likely the omission of vacuum chamber wall effects or the assumption of a single spatially uniform $\alpha_{an}$. \textcolor{black}{It is key to also consider that the experiments source tube walls extended about 0.5$R_c$ into the expansion region beyond the MN throat, which may have acted to impede the free expansion of the plasma. Little and Choueiri also found the largest error to their numerical model to be underestimation of downstream $n_e$ \cite{b:little2016}.}

\textcolor{black}{Although the Bohm model agrees well with the experimental results here, this does not necessarily confirm that it is the definitive or most accurate model for describing anomalous diffusion. However, the strong agreement, mostly within the errorbars of all measurements, does indicate that the general PIC model effectively captures the key physics governing the transport processes, even if the exact mechanisms may differ or be more complex than those assumed in the Bohm formulation. This agreement provides a solid foundation for the controllable preliminary parametric study presented next, allowing for systematic exploration of key transport dependencies while maintaining a physically motivated approach.}

\section{Results}

\subsection{Thruster description and numerical setup}

\subsubsection{150 W MEPT}

The thruster analysed in this work is a particular configuration of the low-power 150 W class MEPT `REGULUS-150', described in references \cite{b:ereg,b:iac2022,b:andrewsiodine}. The cylindrical plasma source has length 60 mm and radius 10 mm. The MN is generated by two rings of permanent magnets which, at the thruster outlet, provide an on-axis strength of $B_0=450$~G.
The thruster was operated, with xenon at $\dot{m}=0.2$~mg/s, at total input powers of $P_{in}=50 - 180$~W. Due to the power drawn by the electronic subsystem, this results in a power delivered to the RF antenna of approximately $P_{RF}=(1-\epsilon_{sys})P_{in}$, where $\epsilon_{sys}$ is the system inefficiency and taken to be about 0.2. The antenna wave-coupling efficiency is estimated to be $\eta_A\approx 0.7$, so the power actually absorbed into the plasma is $P_{a}= (1-\epsilon_{sys})\eta_AP_{in}$. \textcolor{black}{These values for $\epsilon_{sys}$ and $\eta_A$ come directly from measurements made of REGULUS-150 \cite{b:ereg,b:iac2022,b:andrewsiodine}}.
\subsubsection{Global Model}
\begin{figure}[!htb]
\centering
\includegraphics[width=0.7\linewidth]{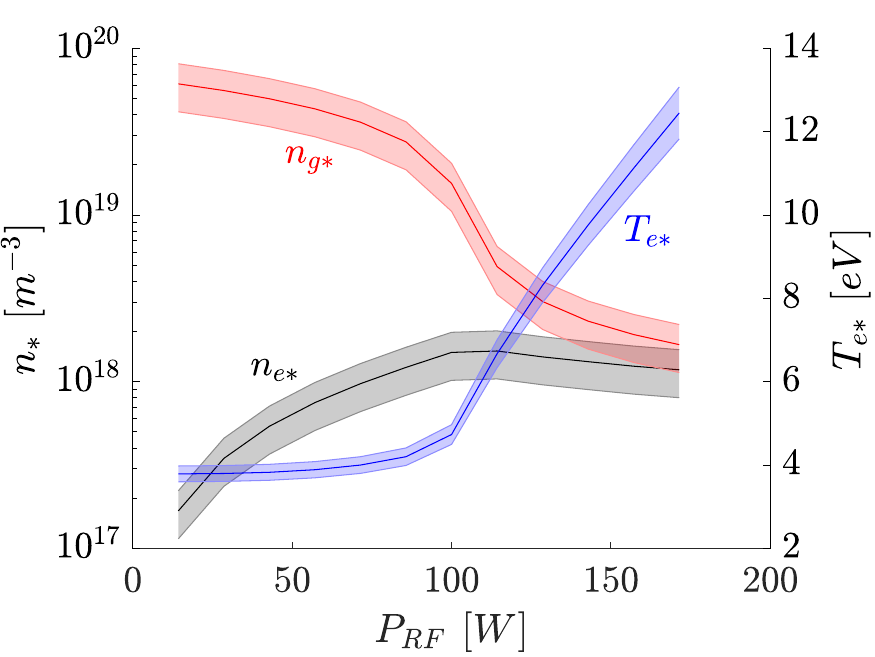}
\caption{\label{fig:GMresults}Discharge properties at the source tube exit from the 0D Global Model: electron density $n_{e*}$ ({\full}), neutral density $n_{g*}$ ({\color{black}\full}), and electron temperature $T_{e*}$ ({\color{black}\full}) }
\end{figure}
The 0D Global Model (see \ref{sec:appendixA}) was used to obtain the properties of the plasma discharge at the source tube exit. The main parameters (i.e. those that serve as inputs to the PIC model) are given in \fref{fig:GMresults} as a function of $P_{RF}$, where the errorbars are a result of the uncertainty in collision cross sections and sheath assumptions described in reference \cite{b:souhairPP}. Two distinct regions of operation are visible. At $<$80 W, the electron temperature and neutral gas density are nearly constant, whilst the electron density steadily increases. At around $100$~W, there is a transition, and the neutral density decreases, with electron temperature rapidly increasing. Electron density remains nearly constant, as the neutral-to-plasma ratio approaches unity.
\begin{figure}[!htb]
\centering
\includegraphics[width=0.7\linewidth]{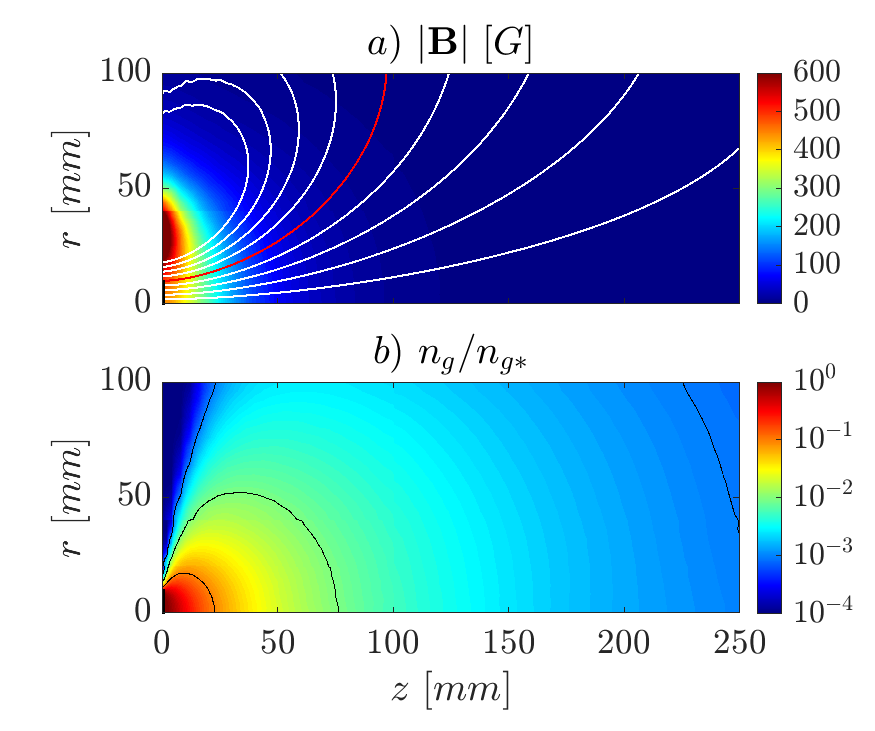}
\caption{\label{fig:bfield}(a) Magnetic field magnitude $|\mathbf{B}|$, the red contour indicates the outermost magnetic field line at the plasma source. (b) Normalised neutral density $n_g/n_{g*}$ from the DSMC simulation.}
\end{figure}
\begin{figure}[!htb]
\centering
\includegraphics[width=1.2\linewidth]{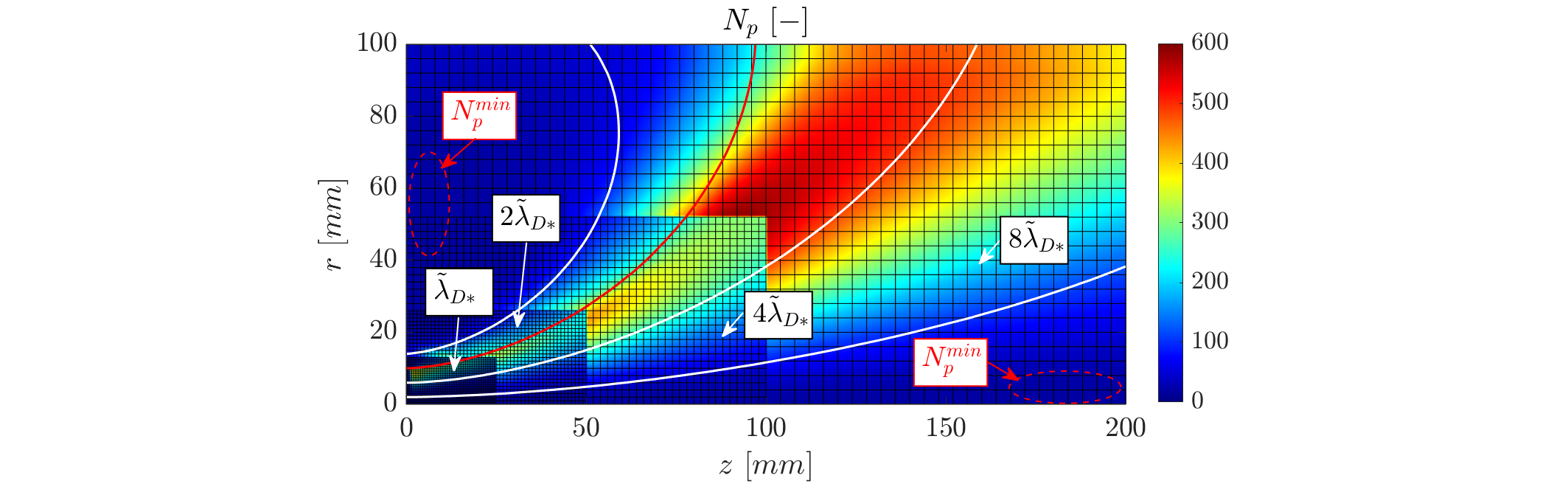}
\caption{\label{fig:np} \textcolor{black}{Piecewise mesh topology, scaled with Debye length, with the distribution of macro-particle number per cell $N_p$ for $P_a=30$ W and $\alpha_{an}=0$.}}
\end{figure}

\subsubsection{PIC model setup}

\begin{table*}[!htb]
\caption{\label{tab:verparams}  \textcolor{black}{Simulation parameters: REGULUS-150-Xe.}}
\lineup
\centering
\begin{indented}
\item[]\begin{tabular}{@{}lllll@{}}
\br
Parameter                          &                               & Unscaled             & Scaled$^a$  & Units            \\ \mr
Thruster Outlet Radius               & $R_0$                  & 10                   & -       & [mm]             \\
Ion Mass (Xe)$^a$                      & $m_i$                  & $131.3$ & {$0.53$} & [AMU] \\
Propellent Mass Flow Rate$^a$          & $\dot{m}$         & 0.25                & 3.95 & [mg/s]                \\
Reference Plasma Density$^b$           & $n_{e*,i*}$              & [$1.8\times 10^{17}-1.2\times 10^{18}$]     & - & [m$^{-3}$]                    \\
Reference Neutral Density$^b$ & $n_{g*}$ & [$1.3\times 10^{18}-6.0\times 10^{19}$] & - & [m$^{-3}$] \\
Reference Electron Temperature$^b$     & $T_{e*}$              & [4-13]                    & -    & [eV]                \\
Reference Ion/Neutral Temperature          & $T_{i*,g*}$              & 298                  & -   & [K]                 \\
Ion Injection Speed$^a$              & $u_{iz0}$        & $c_*$                 & $c_*\sqrt{f}$  & -             \\
Magnet Radius                & $R_c$                   & 25.2                  & -        & [mm]            \\
Throat Magnetic Field Strength     & $B_0$                   & $450$              & -    & [G]                \\ \mr
Axial Domain Length                & $L_z$                  & $20R_0$                 & -      & -              \\
Radial Domain Length               & $L_r$                  & $10R_0$                  & -     & -               \\
Capacitance & $C$ & 0.8 & - & [nF] \\
\br
\end{tabular}
\item[] $^a$ $f=250$.
\item[] $^b$ See Figure 3.
\end{indented}
\end{table*}

\textcolor{black}{The values of $n_{e*}$, $n_{g*}$ and $T_{e*}$ from the 0D Global Model serve as the inputs to the PIC model as described in section 3. The magnetic topology provided by the permanent magnets is illustrated in \ref{fig:bfield} (a) on the simulation domain, which spans $L_z=250$~mm and $L_r=100$~mm. A piece-wise uniform mesh is used, with $\Delta z_0 =  \Delta r_0= R_0/20 = 0.5$~mm at the thruster outlet, with blocks doubling in cell size toward the external boundaries. Its topology is given in \fref{fig:np}. Since $\Delta z \leq \pi\lambda_D$, the value of the artificial permittivity is adjusted to meet this constraint. That is $\gamma=R_0/20\lambda_{D*}$. In this regard, note that it is the ratio of $\tilde{\lambda}_D$ to the system length scale that is kept constant, not $\tilde{\varepsilon}_0$, with the ratio $\tilde{\lambda}_{D*}/R_0=20$ given as per the convergence study in \cite{b:andrews2022}, where the diacritic $\sim$ refers to quantities using the scaled permittivity $\tilde{\varepsilon}_0=\gamma^2\varepsilon_0$ and $\tilde{\lambda}_D=\gamma\lambda_D$}. \textcolor{black}{For the low-power 30 W and high-power 150 W cases concentrated on here, this gives $\gamma \approx 25$ and $17$ respectively}. The associated time-step is constrained by the electron Courant-Friedrich-Lewy condition at $T_{e*}$, thus $\Delta t = 0.5\Delta z_0/3v_{e,th}$. Ions and neutrals have a mass scaling of $f=250$ applied \cite{b:andrews2022}. Since the neutral gas expands from the thruster \textcolor{black}{ only under the influence of gas-dynamic forces}, the neutral density field is self-similar across the operating powers. \textcolor{black}{That is it scales in proportion to $n_{g*}$}. Its normalised value is therefore given in \fref{fig:bfield} (b) as computed by the DSMC module of the code. 

\textcolor{black}{In a study of anomalous transport, it becomes particularly important to eliminate statistical PIC noise ($\propto \sqrt{N_p}^{-1}$) which may produce non-physical numerical diffusion. The average number of macro-particles per cell at the injection plane was ensured to be $\hat{N}_{p0}>50$ per charged species by adjusting the macro-particle weight $W_p$ per simulation case. This remains more than sufficient in accordance with the previous sensitivity study conducted in the previous work of reference \cite{b:andrews2022}. Since the aforementioned piece-wise uniform mesh is scaled according to the local Debye length cell-size criterion, the variance in $N_p$ is kept small and this is illustrated in \fref{fig:np} for the example of the $P_a=30$ W, $\alpha_{an}=0$ case; A minimum value of $N_p^{min}=21$ per species was seen in the region of the upstream radial boundary and downstream symmetry axis. In an axisymmetric domain, the cell volume scales with radius due to the cylindrical coordinate system. Near the symmetry axis at $r=0$, the annular cell volumes become very small, leading to fewer macro-particles per cell despite the more-uniform particle density in physical space. Larger cell volumes at larger radius, combined with the MN confinement, results in greater $N_p$ just inside the outermost magnetic filed line connected to the source, with $N_p^{max}=590$. This gives a globally average number of macro-particles per cell as $\hat{N}_p=200$ per species. With $W_p$ managed such that $\hat{N}_{p0}>50$ in each case, the values of $N_p^{min}$ and $\hat{N}_p$ remain similar across simulations.} Finally, the virtual capacitance was set to $C=0.8$~nF throughout.

Two parameters are varied in the simulations here. Firstly the value of the absorbed power, has been investigated for 30 - 150 W. This is equivalent to altering the neutral-to-plasma density and electron temperature according to \fref{fig:GMresults}.  Then, the effect of the anomalous transport is analysed by increasing the empirical Bohm parameter $\alpha_{an}$ from 0 (classical) to 1/16 in the fully-turbulent limit. \textcolor{black}{The magnitude of the imposed anomalous collisionality is also first compared to the classical in section 4.2.}. \textcolor{black}{All results presented have been time-averaged for 50000 time-steps after steady-state was reached}.

\subsection{Classical and anomalous collisionality}
\begin{figure*}[!htb]
\includegraphics[width=\linewidth]{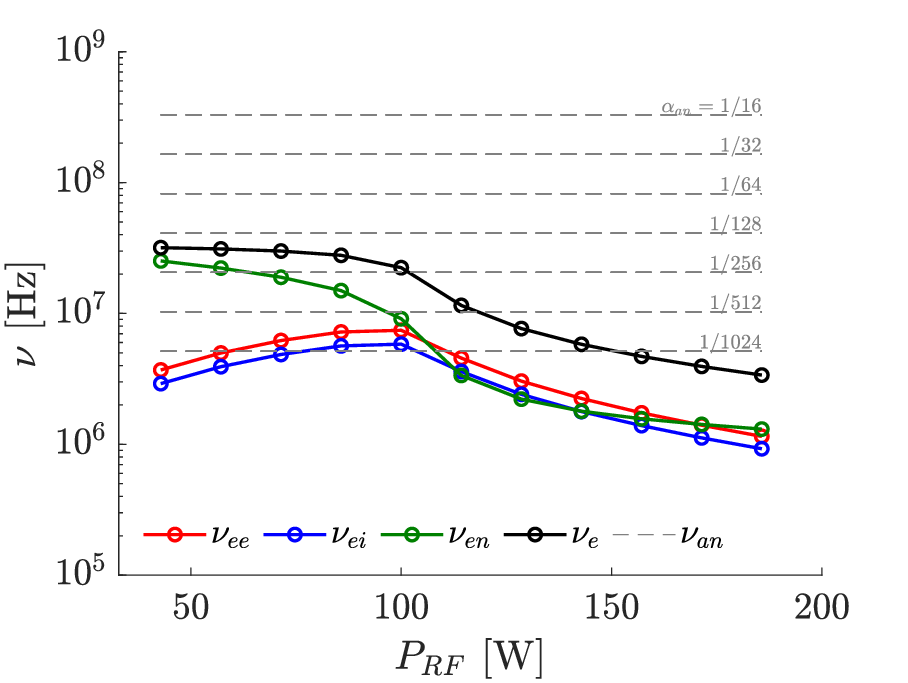}
\caption{\label{fig:collfreq} Comparison of the classical electron collision frequency components, as a function of RF power, to the imposed magnitudes of anomalous Bohm collisionality.}
\end{figure*}
\textcolor{black}{It is prudent to first contextualize the magnitude of the imposed Bohm collisionality relative to classical collisionality to better grasp its significance. This comparison provides a baseline for estimating the extent of anomalous transport and offers insight into deviations from classical expectations. For each classical simulation case at steady-state, the average electron collision frequency components, within the first $R_0\times R_0$ region of the domain, were recorded directly from the MCC module of the code. \Fref{fig:collfreq} presents these collision frequencies as a function of the thruster input power $P_{RF}$, where the increasing values of the Bohm coefficient are indicated as reference levels for comparison.}

\textcolor{black}{As thruster power increases from low to moderate levels, both electron-electron and electron-ion Coulomb collision frequencies initially rise from circa $3\times 10^{6}$ Hz to $6\times 10^6$ Hz. Higher power input leads to greater mass utilisation efficiency, which increases plasma density, thereby enhancing the likelihood of Coulomb collisions. However, beyond 100 W, further power increase leads to a significant rise in electron temperature, which reduces Coulomb collision frequency to around $1\times 10^6$ Hz at 185 W. This occurs because higher-energy electrons have reduced Coulomb collision cross-section (proportional to $T_e^{-3/2}$) and due to the electron velocity dependence of the Coulomb logarithm. }

\textcolor{black}{The electron-neutral collision frequency, which includes both elastic scattering as well as inelastic ionisation and excitation components, is strongly a function of the neutral gas density. It dominates the classical transport at low-power by an order of magnitude; around $2.5\times 10^7$ Hz at 30 W. With increasing power, the neutral density quickly decreases, thus so to does the electron-neutral collision frequency. Beyond 110 W, it is of similar magnitude to Coulomb collisions. At high-power, it plateaus as, with rapidly rising average electron temperatures exceeding the excitation and ionisation thresholds, the inelastic collision frequencies increase.} 

\textcolor{black}{The total classical electron collision frequency is therefore on the order of $10^6-10^7$ Hz: At low-power (30-85 W), $\nu_e$ is consistent at around $3\times 10^7$ Hz before undergoing a steady transition beyond 100 W to a near-linear decline. At 185 W, $\nu_e=3.3\times10^6$ Hz; therefore, there is an order of magnitude drop in collisionality (and cross-field transport) between low and high-power. Comparing this to the graduations of the Bohm scaling on \fref{fig:collfreq}, it can be seen that, while a value of $\alpha_{an}=1/1024$ is sufficient to dominate the transport at high-power, values $\alpha_{an}\geq 1/128$ are required to exceed the classical collision frequency below 115 W. It can be expected then that there will exist a critical value of $\alpha_{an}$, per operating condition (power), that will yield an anomalous diffusion significant enough to supersede the classical behaviour. In the fully-turbulent limit of $\alpha_{an}=1/16$ ($\nu_{an}=3.2\times 10^9$ Hz), anomalous collisions dominate by one and two orders of magnitude relative to the classical collisions at low and high-power respectively.}

\subsection{2D plasma profiles}
\begin{figure*}[!htb]
\includegraphics[width=\linewidth]{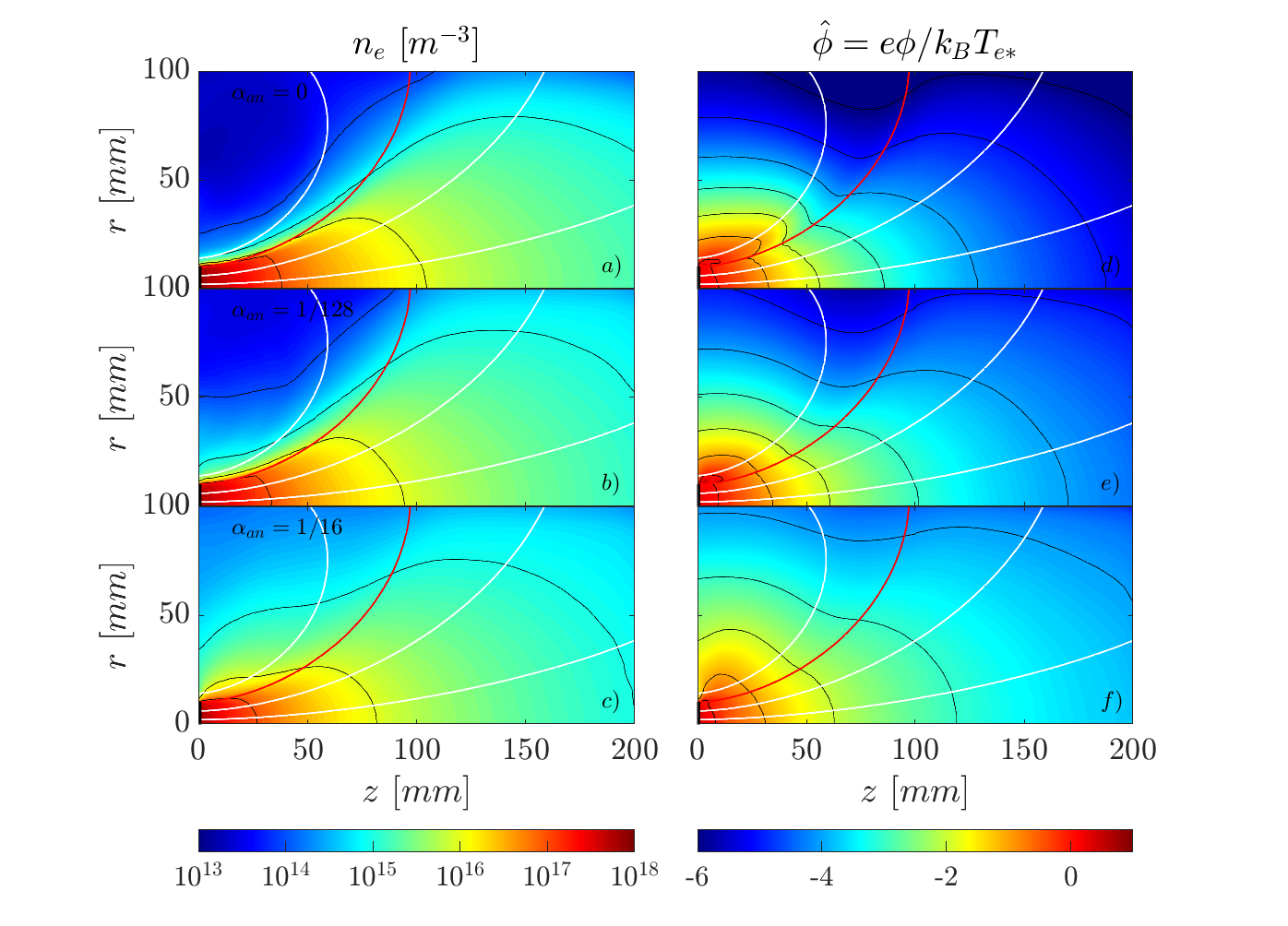}
\caption{\label{fig:2Dprofiles} 2D maps of number density $n_e$ and normalised plasma potential $\hat{\phi}=e\phi/k_BT_{e*}$ at $P_{a}=30$~W ($n_{e*,i*}=5.41\times 10^{17}$ m$^{-3}$, $n_{g*}=4.99\times 10^{19}$ m$^{-3}$, $T_{e*}=3.83$ eV) and for $\alpha_{an}=0$, $1/128$, and $1/16$.}
\end{figure*}

\textcolor{black}{\Fref{fig:2Dprofiles} presents the 2D maps of the electron density $n_e$ and normalised plasma potential $\hat{\phi}=e\phi /k_B T_e*$ for the $P_{a}=30$~W ($n_{e*,i*}=5.41\times 10^{17}$ m$^{-3}$, $n_{g*}=4.99\times 10^{19}$ m$^{-3}$, $T_{e*}=3.83$ eV) case at three levels of the anomalous parameter $\alpha_{an}$: $0$, $1/128$ and $1/16$. Fundamentally, in the axial direction, the plume propagates monotonically downstream under the dominance of the self-consistently developed ambipolar electric field. In the classical case ($\alpha_{an}=0$), the plume is highly-collimated within the boundary of the outermost magnetic field line (OMFL) connected to the plasma source. The density exhibits canonical magnetic confinement characteristics with pronounced isocontours approximately parallel to magnetic field lines in the near plume, and steep density gradients perpendicular. The electron transport is predominantly governed by the electron gyro-advection with minimal cross-field transport from classical collisions. As anomalous transport intensifies at $\alpha_{an}=128$, the cross-field electron mobility increases proportionally to $B^{-1}$ and in excess of the total classical collision frequency (see Figure 7). This manifests as a discernible deviation of the $n_e$ isocontours from the magnetic field lines, particularly in regions of high magnetic field gradients. The perpendicular density scale length increases significantly compared to the classical case, indicating the expected relaxation of cross-field density gradients while still maintaining the general magnetic field-aligned structure. At $\alpha_{an}=1/16$, the collimated plume structure is lost. The cross-field diffusivity gives substantial departure from classical magnetic confinement within the OMFL, resulting in a much broader radial plasma distribution. The density isocontours show marked deviation from magnetic field lines; There is quasi-isotropisation of the density gradients in regions where the magnetic field is not sufficiently strong to counteract the anomalous transport by gyro-advection. This is particularly evident in the plume periphery beyond the OMFL, where the magnetic field strength diminishes, resulting in a quasi-radial diffusion-dominated regime rather than the magnetically-guided regime observed at lower $\alpha_{an}$ values. The increased cross-field mobility facilitates electron transport into previously inaccessible regions of the domain, effectively broadening the plasma expansion and reducing axial density gradients. This progressive transition from magnetically-dominated to diffusion-influenced electron transport has profound implications for plume dynamics and energy transfer mechanisms in the MN. The change in density distribution alters the electron pressure gradient forces that contribute to the azimuthal current formation and thrust generation, discussed later in this section.}

\textcolor{black}{Concerning the plasma potential of \fref{fig:2Dprofiles}s (d-f), a strong ion-confining potential barrier, indicated by the non-monotonic radial discontinuity in isopotential, forms along the OMFL. This structure has been confirmed in many other numerical works \cite{b:chen2020,b:chen2022,b:andrews2022,ionpeak}. The potential barrier represents a critical self-organising mechanism for ion confinement, forming in response to the radial ion energy flux, since the local electric field from the barrier returns the secondarily expanding ion trajectories back into the MN. It also enhances the transport of electrons towards this magnetically isolated region. Both the location and strength of the potential barrier is therefore highly dependant on ion injection properties including temperature, Mach number, or pre-injection divergence angle \cite{b:chen2020,ionpeak}. Local conditions, such as ion-neutral charge-exchange collisions \cite{b:andrews2022} and thruster casing boundary conditions on the electric field (conducting, charge-accumulating dielectric etc.) \cite{b:andrewsiodine}, also act to determine the effective radial ion energy flux. Previous work on the low-power 50 W variant of the REGULUS thruster has shown much stronger potential barriers than the 150 W class thruster considered here, primarily due to increased charge-exchange \cite{b:nabiliepc,b:andrews2022}.}

\textcolor{black}{In the classical case ($\alpha_{an}=0$), the potential barrier is well-defined along the OMFL for $z<60$ mm as the differential ion and electron confinement creates substantial ambipolar perpendicular electric fields, manifesting a peak potential of 0.89$T_{e*}$. With increasing $\alpha_{an}$, electrons are able to respond more directly to cross-field gradients, due to increased perpendicular mobility, rather than being constrained to $E\times B$ gyro-drift along equipotential surfaces. Electron cross-field transport increasingly enables their ambipolar coupling to the unmagnetised ion expansion, thereby reducing the charge separation that generates the potential barrier. At $\alpha_{an}=1/128$, the potential barrier is significantly weaker, exhibiting diminished steepness of the electric field along the OMFL and a peak of 0.70$T_{e*}$. The location of the potential barrier begins separating above the OMFL at $z\approx 40$ mm but, in the radial plume far-field, the isopotentials are near-identical in profile to those of the classical case. In the far-plume, the Bohm diffusion ($\propto 1/B$) is small and thus the magnetic advection once again dominates the electron transport, allowing full formation of the potential barrier. At $\alpha_{an}=1/16$, the potential barrier in the near-plume is no longer distinguishable with near-homogeneous isopotential distribution from the plasma source. A path of high radial potential, similar to that formed in an unmagnetised case is seen, as the plume follows a \textcolor{black}{more ambipolar} expansion. This allows the radial flow of ions required by plasma quasineutrality. $At z> 50$ mm, a weak discontinuity in the isopotentials begins to recover, separated considerably from the OMFL. It will be shown later that the structure of the potential barrier can significantly affect the distributions of the azimuthal electron currents which generate magnetic thrust.}

\textcolor{black}{Notable in \fref{fig:2Dprofiles}s (d-f) is also the reduced potential fall, both axially and radially, with increasing $\alpha_{an}$. From source to infinity, the global potential drop at $\alpha_{an}=0$ is $|\phi_\infty|=31.8~V$, about $8.29T_{e*}$. The drop lessens significantly as $\alpha_{an}$ increases, as given in \tref{tab:drop}, which includes results for additional values of $\alpha_{an}$. The potential drop approaches the unmagnetised theoretical $\phi_\infty/T_{e*}=0.5(1+\ln{{m_i/2\pi m_e}})\approx 5.77$, given per Lafleur \cite{b:lafleur2014}, toward the fully turbulent case. A sharp decline between $\alpha_{an}=1/256$ and 1/128 represents the point at which Bohm collisions begin to greatly exceed the classic collisionality as per \fref{fig:collfreq}. Anomalous diffusion reduces the potential drop because it dominates the magnetic advection that ordinarily restricts electron transport pathways in their velocity phase space. Cross-field transport widens the electron loss cone in velocity space, allowing electrons with lower parallel velocities to reach the infinity through perpendicular transport. This lowers the average electron energy escaping the MN. In addition, significant cross-field diffusion in the near-plume reduces the magnetic mirror reflection of those electrons returning to the plasma source.  The plasma no longer needs to develop as large a downstream potential drop to maintain quasi-neutrality and current balance with the outflowing ions. The electron velocity distribution requires less severe truncation in the parallel direction, as more electrons across the distribution can contribute to the current to infinity via cross-field transport. This directly impacts the final ion acceleration (and thrust).}

\begin{table}[!htb]
\centering
\caption{Potential drop at $P_{a}=30$~W for increasing $\alpha_{an}$}
\label{tab:drop}
\begin{tabular}{@{}lllllll@{}}
\toprule
$\alpha_{an}$             & 0 & 1/256 & 1/128 & 1/64 & 1/32 & 1/16 \\ \midrule
$e|\phi_\infty|/k_BT_{e*}$ & 8.29 & 8.21 & 6.78 & 6.57 & 6.22 & 5.96    \\ \bottomrule
\end{tabular}
\end{table}
\begin{figure*}[!htb]

\includegraphics[width=1\linewidth]{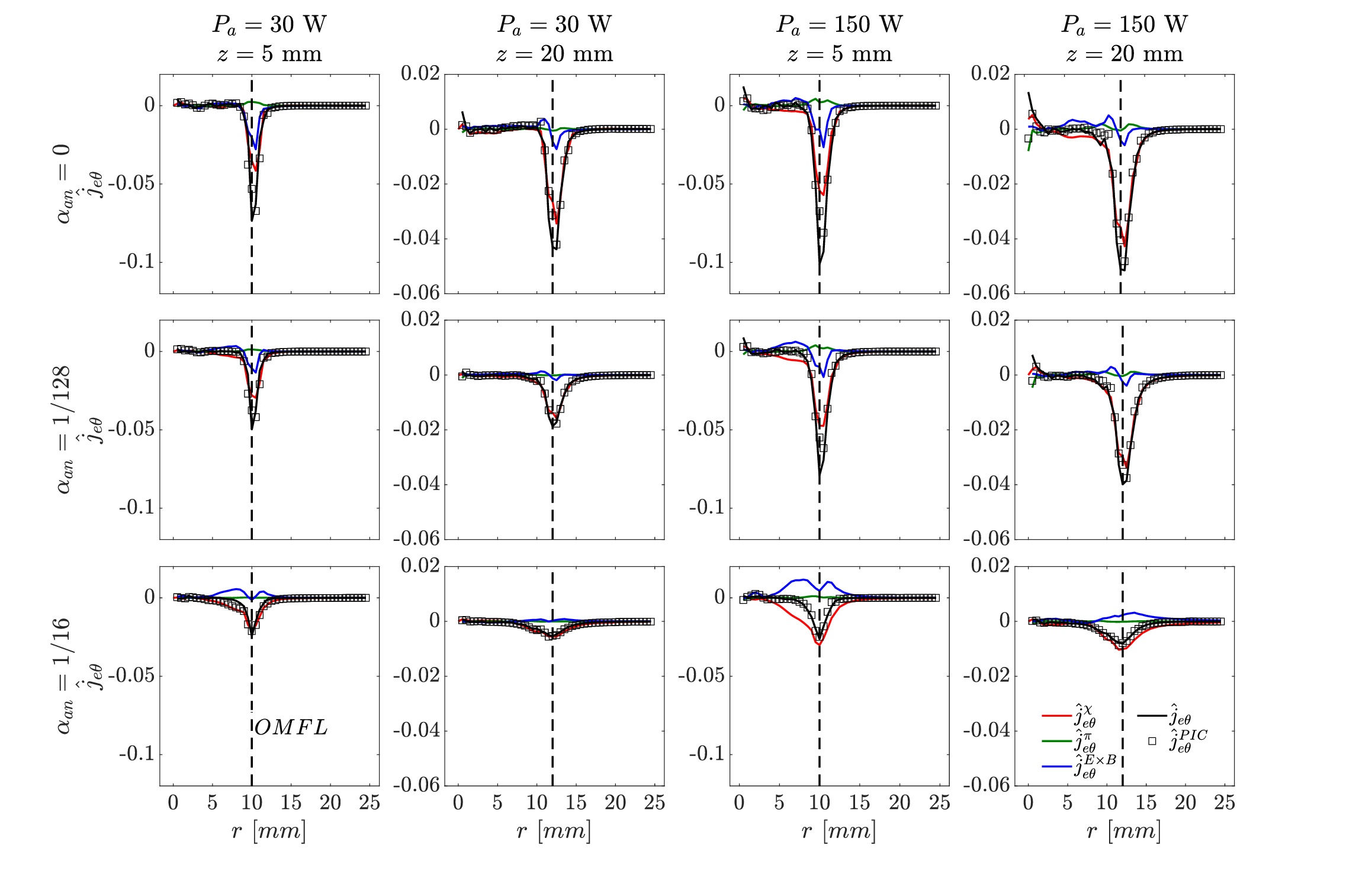}
\caption{\label{fig:azcurrent} Radial azimuthal current density profiles for $\alpha_{an}=0, 1/128$ and $1/16$ at $P_{a}=30$ and 150~W for $z = 5$ and 20 mm. The vertical dashed line indicates the location of the OMFL.}
\end{figure*}
\begin{figure}[!htb]
\includegraphics[width=\linewidth]{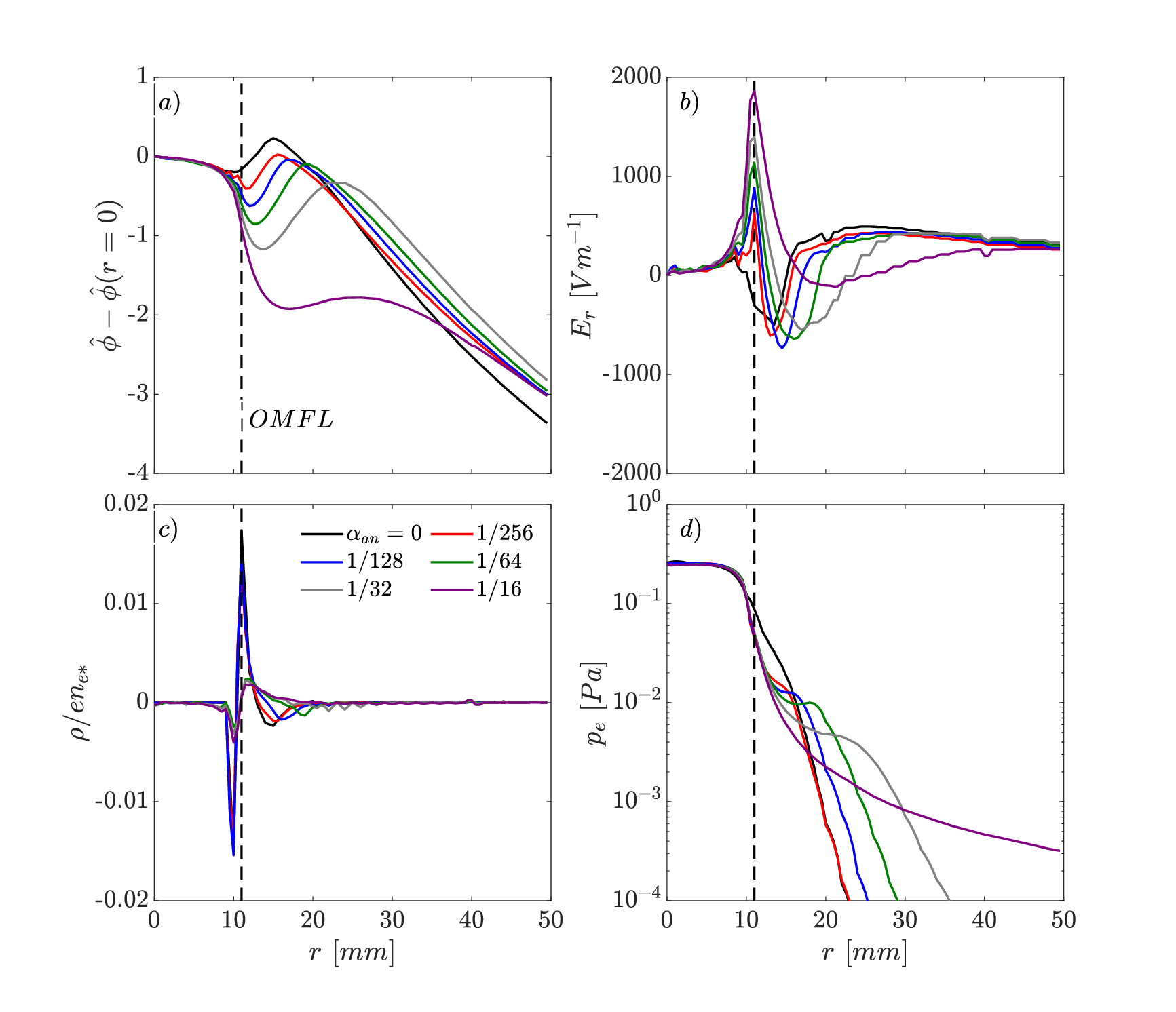}
\caption{\label{fig:radialalpha} A comparison between the (a) radial plasma potential profiles, (b) radial electric field, and (c) charge density for increasing values of $\alpha_{an}$ in the $P_{a}=30$~W case \textcolor{black}{at $z=5$ mm}.}
\end{figure}
\begin{figure*}[!htb]
\includegraphics[width=\linewidth]{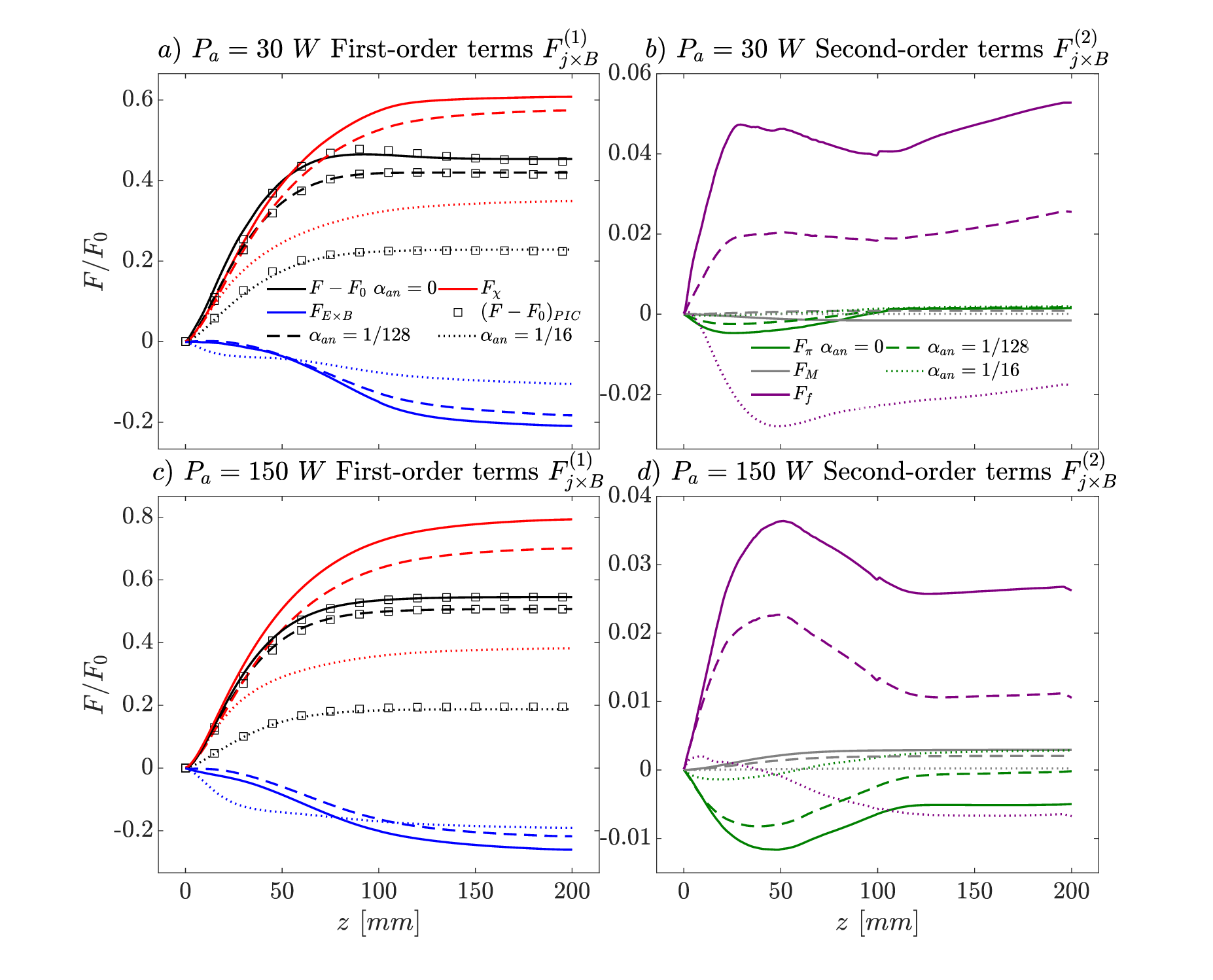}
\caption{\label{fig:thrustaxis} Electromagnetic thrust components accumulating along the MN for 30 W (a-b) and 150 W (c-d), split by first and second-order components $F_{j\times B}=F_{j\times B}^{(1)}+F_{j\times B}^{(2)}$.}
\end{figure*}
%
\subsection{Azimuthal current formation and thrust generation}

\Fref{fig:azcurrent} shows the radial profiles of azimuthal current densities at $z= 5$ and 20~mm for both $P_a=30$ and 150 W, in which the contributions were calculated as per equation \eref{eqn:azimuthalcurrent}; the current densities are normalised by $en_{e*}v_{th,e*}$ (the injected thermal current density). The radial position of the OMFL has also been marked as a vertical dashed line for reference. For each case, the sums of all current density contributions $\hat{j}_{e\theta}$ fit well (error below 3\%) with the directly calculated PIC total current density $\hat{j}_{e\theta}^{PIC}$. The friction-induced and inertial currents were found to be small in all cases $j_{e\theta}^{f,M}/{j}_{e\theta}=\mathcal{O}(10^{-2})$ and thus are omitted for clarity.

At both powers, the magnitude of diamagnetic flow current $\hat{j}_{e\theta}^{\chi}$ decreases with increasing $\alpha_{an}$. The maximum intensity of diamagnetic current corresponds to the OMFL and, at 5 mm, decreases from $-4\times 10^{-2}$ to $-1.2\times 10^{-2}$ at 30 W and from $-5\times 10^{-2}$ to $-3\times 10^{-2}$ at 150 W. At 30 W, this is highly focused around the OMFL; at 150 W, the distribution is much broader, with significant current density $\pm 2$ mm from the OMFL. The same observations are seen at 20 mm, but the intensity is approximately 60-70\% that at 5 mm. This trend is because, at higher $\alpha_{an}$, there is a weaker plasma confinement along the OMFL, reducing the electron pressure gradient $\partial p_e/\partial r$. \textcolor{black}{It is judicious to clarify here that, in reality, the steady-state pressure gradient will self-consistently result from the complex interplay between various drift-driven instabilities and particularly the contributions of additional drifts that influence the saturation of these instabilities. This is not captured in the phenomenological approach here}.

At 0 and 1/128, the $\mathbf{E} \times \mathbf{B}$ drift currents $\hat{j}_{e\theta}^{E \times B}$ have both a paramagnetic---prior to the OMFL---and diamagnetic contribution---on the OMFL. This is due to aforementioned potential barrier; the plasma potential does not drop monotonically in the radial direction. Rather, it decreases until the OMFL, and rises again to form the potential barrier outside the plume periphery. At 1/16 however, $\hat{j}_{e\theta}^{E \times B}$ is always paramagnetic. The explanation for this is that the reduced magnetisation leads to the loss of the potential barrier and a higher hall angle (i.e. the angle between the electron current and the magnetic field). \textcolor{black}{Since the expansion approaches that of an unmagnetized plasma as the cross-field diffusion is high}, the electric field is less-aligned with the MN, and thus $j_{e\theta}^{E\times B}$ is increased. The paramagnetic current density at 150 W peaks at $1.2\times 10^{-2}$ at 5 mm compared to $5.0\times 10^{-3}$ at 30 W. This is because the source ions have more energy at higher power (and a number of low-energy ions from in-plume ionisation will be accelerated radially), yielding a greater radial electric field to counter increased ion secondary expansion. 

As expected the stress-induced current is small $j_{e\theta}^{\pi}/{j}_{e\theta}=\mathcal{O}(10^{-1})$ \cite{b:Hu2021}; at 30 W and $\alpha_{an}=0$, it has a $1.3\times 10^{-3}$ maximum at the OMFL at 5 mm, decreasing to negligible at 1/16. \textcolor{black}{As electrons gyrate about magnetic field lines, they exchange momentum between adjacent plasma layers with different macroscopic flow speeds, generating a gyrotropic (anisotropic in $\mathbf{B}$) shear stress analogous to a gyration-induced viscosity. Due to the fast electron gyromotion, this results in a resistance to rapid changes in flow direction. As a result, faster-moving layers are slowed down, while slower layers are pulled along. This stress transfers momentum between the layers, reducing velocity shear and smoothing out sharp flow gradients.} In the fully-turbulent case, anomalous collisions act to thermalise the electrons toward an isotropic population, therefore eliminating the viscosity. At 150 W, $j_{e\theta}^{\pi}$ is generally about 4-times greater than at 30 W, owing to the smaller classical collisionality and so increased effects of electrons drifting radially off the magnetic field lines. Importantly, at 5 mm, $j_{e\theta}^{\pi}$ is paramagnetic, \textcolor{black}{which is detrimental to magnetic thrust generation}. But, at 20 mm, it becomes diamagnetic. When the electrons start to demagnetise as the magnetic field drops in the downstream, the cyclotron radii of electrons becomes comparable to the system scale and there is an acceleration from the finite electron Larmor radius effect \cite{b:chen2022}.

Further insight is found in \fref{fig:radialalpha}s (a-d) for six different values of the anomalous parameter $\alpha_{an}$, where the radial slice at $z=5$~mm is taken since this is where the plasma potential peak is greatest. It can be seen in \fref{fig:radialalpha}(b) that the potential well coincides with a direction reversal of the radial electric field near the edge of the plasma expansion. In the classical theory, ions with sufficient radial energy near the MN periphery are able to overshoot the edge of the attached electron fluid, and create an excess of positive charge just beyond the boundary, clear in \fref{fig:radialalpha}(c), with a comparable negative charge layer just within the boundary. 
As $\alpha_{an}$ increases however, electrons are increasingly able to diffuse across the MN periphery with the ions. This cross-field electron transport balances some of the space-charge and hence dampens the electric field across the OMFL. A potential well of sufficient depth to return the secondary-expanding ions towards the MN can no longer form, and the confinement is greatly diminished as a result. An interesting result of \fref{fig:radialalpha}(c), is a defined transition in the charge density profiles between $\alpha_{an}=1/128$ and $1/32$. This agrees well with established observations in the literature that there exists a critical condition where a magnetised expansion is collimated, or otherwise under-collimated.

The resultant impact on thrust relies on the volumetric integral of equation \eref{eqn:force}. By axially truncating the control volume \cite{b:andrews2022}, the axial evolution along the MN of the magnetic thrust components are given in \fref{fig:thrustaxis} for both $P_{a} =$ 30 W and 150 W. \textcolor{black}{Also presented is the thrust calculated directly from individual particle momentum $(F-F_0)_{PIC}$, where the error is seen to be $<3$ \% to that from the sum of the fluid moments of the PIC distribution.} Generally, the diamagnetic flow effect is an accelerating contribution, while $E\times B$ drift induced can be classified as decelerating ones \cite{b:ahedo2010}. In all cases, the diamagnetic effect is the dominating term, covering over 90\% of positive contribution in each region. The diamagnetic-flow current produced by $\mathbf{\nabla}p_e\times\mathbf{B}$ interacts with the applied field, generating a propulsive Lorentz force. In this way, the electron internal energy is converted into the axial kinetic energy of ions. As seen in \fref{fig:thrustaxis}s (a) and (c), the diamagnetic thrust is inhibited by the anomalous collisionality by approximately 5\% and 40\% for 1/128 and 1/16 respectively. Increased electron diffusion results in a lower $\partial p_e / \partial \mathbf{1}_\perp$, observed in \fref{fig:radialalpha} (d), by weakening the confinement at the OMFL, allowing more plasma to expand beyond the MN. The more-weakly magnetised nozzle therefore has a detachment point shifted upstream, as is clear in the earlier convergence of $F_{\chi}$ in the cases of anomalous diffusion.

The \textcolor{black}{induced} $E\times B$ term is the major component of deceleration. The reason for this is the large area of paramagnetic induced currents caused by the inward detachment of ions, determined by the radial electric field that acts in response to the charge separation with the gyrating electrons. At 1/16, $F_{E\times B}$ is reduced by 70\% and 50\% for 30 and 150 W respectively in \fref{fig:thrustaxis} (a) and (c). This is since the more weakly confined expansion allows the electron fluid to follow the unmagnetised ions, reducing the magnitude of the radial electric field across the OMFL. An interesting observation however, is that for $\alpha_{an} = 1/16$, this decelerating thrust accumulates rapidly within approximately $50$~mm compared to the slow increase in the classical and 1/128 cases. Although the decelerating effect of $F_{E\times B}$ is reduced by anomalous effects, it does so less than the comparative reduction in the accelerating $F_{\chi}$; the net thrust is reduced by approximately 47\% and 62\% for 30 and 150 W respectively.

The second-order mechanisms are the stress-induced, inertial and frictional components, which exhibit complex behaviour in \fref{fig:thrustaxis} (b) and (d). First, it is seen that the inertia effect, which converts the directed kinetic energy of electrons to that of the ions, is expectantly small ($F_M<5\times 10^{-3}F_0$) due to the tiny mass possessed by electrons. At 30 W, $F_\pi$ is initially decelerating, but tends to a negligible accelerating contribution at about $z=$ 100 mm. At 150 W however, $F_\pi=-5\times 10^{-2}F_0$ in the classical case, representing the neoclassical transport that takes place in this less-collisional case. At 1/16 though, $F_\pi=2.9\times 10^{-3}F_0$, providing a small positive contribution. Note that the axial distribution of the stress component is insensitive to the anomalous diffusion, only its magnitude. Finally, $F_f$ is the friction-induced force; it is dominated by the anomalous axial and radial collisionalities, which were found to be of order $10^9-10^{10}$ Hz and in agreement with the frequencies reported in reference \cite{b:alvaro2021}. In the classical case, $F_f\approx 0.02-0.05F_0$ for both powers, so there is a non-negligible transfer of azimuthal momentum to the electrons from collisions and the anomalous axial-radial instabilities. As $\alpha_{an}$ increases to 1/16, $F_f\approx -10^{-2}F_0$ and becomes a detrimental contribution. Increased electron diffusion in the plume reduces electron axial drift velocity and increases the radial velocity, with a consequent effect on collisions occurring and $F_{f}$.

\subsection{Propulsive performance}

\begin{figure}[!htb]
\centering
\includegraphics[width=0.7\linewidth]{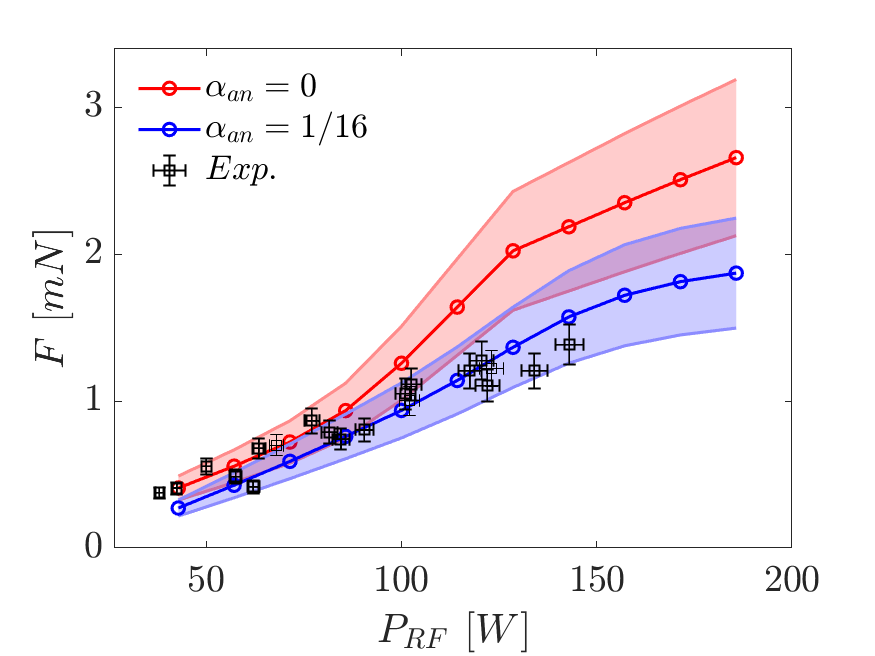}
\caption{\label{fig:thrust} Thrust for $\alpha_{an}=0$ and $\alpha_{an}=1/16$ against experimental measures of REGULUS-150-Xe.}
\end{figure}
Now that the effect of anomalous diffusion on the thrust generation has been concluded in the previous subsection, the scope of the analysis here is to compare, from a macroscopic perspective, the performance with and without anomalous diffusion, and to verify if assuming a fixed value of $\alpha_{an}$ is sufficient to reproduce experimental trends when $P_a$ is varied. A thrust balance, specifically designed for RF thrusters of small-to-medium size was employed as described in \cite{b:trezzolani2019}. The thrust given by the PIC model is compared to these experimental measures; the results are presented in \fref{fig:thrust} for both the 1/128 and 1/16 cases. There is good agreement in the 1/128 case for $P_{RF}<80$ W, but at higher powers the thrust is overestimated by up to 48\%. This is corrected when anomalous collisions are included, with the experimental points lying within the numerical error. However, at lower powers the 1/128 cases retain the better fit. This raises questions regarding the dependence of $\alpha_{an}$ on $P_{RF}$ (or more directly $P_*$). Indeed, the azimuthal fluctuations responsible for anomalous diffusion have been well-documented to be related to the plasma wave energy \cite{b:exb}. This wave energy should be proportional to $P_*$, thus it is possible that the Bohm coefficient is an increasing function of the thruster power. In addition, a non-negligible background facility pressure may suppress some of the anomalous instability which is presently un-characterised \cite{b:andri}; \textcolor{black}{Indeed, preliminary work regarding a self-consistent anomalous diffusion model, derived from lower-hybrid drift-dissipative instability theory, has shown that electron-heavy species collisions can reduce the frequency bandwidth and subsequent anomalous cross-field transport in proportion to $\nu_{e}u_{e\theta}^3$ \cite{b:shauniepc}}. Regardless, the experimental agreement is within 20\%.

\begin{figure*}[!htb]
\includegraphics[width=\linewidth]{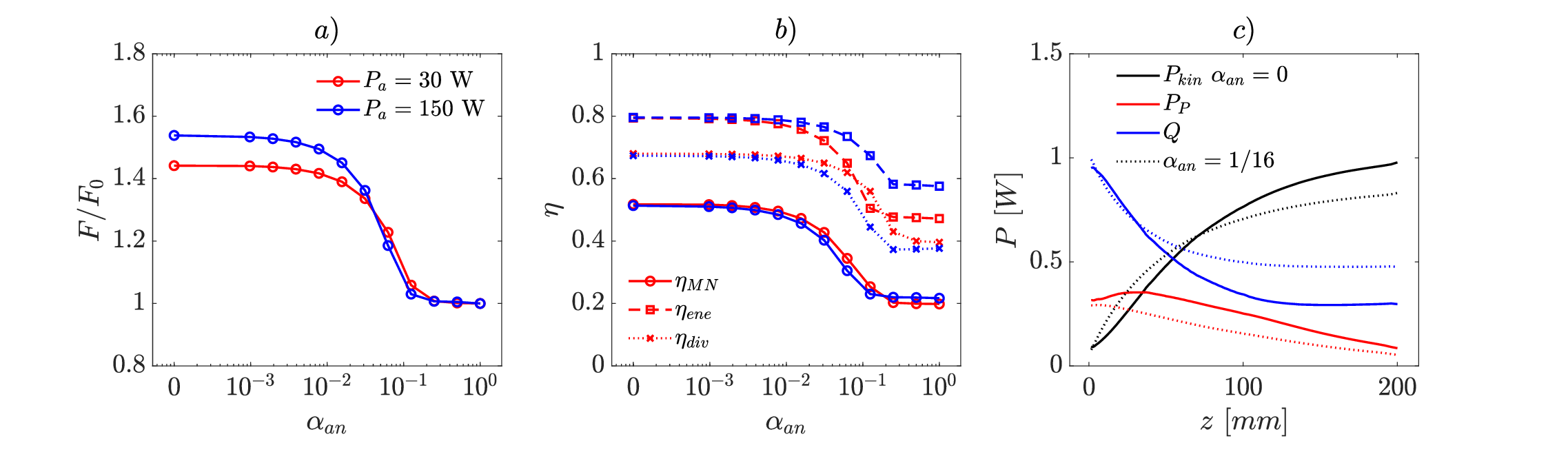}
\caption{\label{fig:bohmperf} Effect of $\alpha_{an}$ on performance: (a) Thrust gain;  (b) Energy, divergence and MN efficiency; (c) Axial evolution of the MN power at $P_a=30~W$.}
\end{figure*}
Further simulations for values of $\alpha_{an}$ between 0 and 1 were carried out, as shown in \fref{fig:bohmperf} (a) for the MN thrust gain. The anomalous transport becomes negligible (compared with the classical one) for \textcolor{black}{ $\alpha_{an}=\nu_e/\omega_{ce} \lesssim \mathcal{O}(10^{-3})$}, with typical values of $\nu_e$ about $10^6 - 10^7~ s^{-1}$ and $\omega_{ce}$ about $10^9 -10^{10}~s^{-1}$. The thrust gain diminishes significantly: For 30 and 150 W respectively, $F/F_0$ decreases from 1.44 and 1.54, when $\alpha_{an}\lesssim 10^{-2}$, towards unity as $\alpha_{an}\gtrsim 10^{-1}$. The plume becomes effectively unmagnetised as the anomalous cross-field diffusion inhibits the field-aligned advection of electrons; they can no longer establish sustained gyrations and thus azimuthal current---and therefore magnetic thrust generation---ceases. There is a critical value of $\alpha_{an}$ around 10$^{-2}$ where the thrust loss rapidly increases, and this occurs at lower $\alpha_{an}$ for 150 W; the demagnetised state also occurs earlier for 150 W, at $\alpha_{an}\approx 1/8$ (as opposed to $\alpha_{an}\approx 1/4$ at 30 W).
To examine the performance impact in more detail, \fref{fig:bohmperf} (b) presents the relevant efficiencies. At both 30 and 150 W, the overall MN efficiency $\eta_{MN}$ decreases from 0.51, aligned with typical values found in other MN simulations \cite{b:chen2022,b:alvaro2021}, to about 0.2. At 150 W, the energy efficiency falls from 0.75 to 0.71 approaching $\alpha_{an}=1/32$, then undergoes a significant but steady drop to 0.58 for $\alpha_{an}\gtrsim 1/4$. The 30 W case follows a similar trend, but experiences a larger relative loss; $\eta_{ene}$ drops from 0.73 to 0.47. As discussed in section 4.3, the demagnetisation of electrons reduces the ambipolar potential drop that facilitates energy conversion in the MN. The divergence efficiency $\eta_{div}$ is around 0.64 for both powers at $\alpha_{an}\lesssim 10^{-2}$, corresponding to an average beam divergence angle of $cos^{-1} \sqrt{\eta_{div}}\approx$ 37$^\circ$. There is then a very rapid decline toward 0.4 (51$^\circ$) for $\alpha_{an}\rightarrow1$, as the magnetic confinement is depleted and the plume approaches unmagnetised ambipolar expansion; the transition to this unmagnetised divergence occurs at $\alpha_{an}$ of 1/8 and 1/2 for 150 and 30 W respectively.

\textcolor{black}{The axial evolution of the power balance, presented in section 2.2.2. and used to establish the efficiency metrics, is given in \fref{fig:bohmperf} for the $P_a=30$ W case, where $P_*=1.36$ W. While the power from thermal pressure remains a small contribution in both the classical and $\alpha_{an}=1/16$ case, it is further confirmed that the performance loss is primarily due to reduced (ion) bulk kinetic power. It is cut from 0.98 to 0.83 W of which only 0.66 and 0.33 W is in the axial direction, per the values of $\eta_{div}$; a 50\% loss. Not revealed in the efficiency metrics is the behaviour of the heat flux power $Q$. \Fref{fig:bohmperf} reveals that the heat flux remains 60 \% higher in the case of anomalous diffusion. This indicates that more energy remains in fast-moving electrons or non-equilibrium electron velocity distributions and thus partially why the MN is not efficiently converting thermal energy into bulk kinetic energy. With increased electron conduction from anomalous collisions, heat flux can remain high if temperature gradients exists downstream. Fast-moving electrons retain their energy longer and thus not transferring it, via the MN, to propulsive kinetic energy. Note that in the 30 W case, the electron temperature is too low to result in any significant $P_{loss}$ from inelastic collisions.}

\section{Insight into anomalous electron cooling}

 \begin{figure*}[!htb]
 \includegraphics[width=\linewidth]{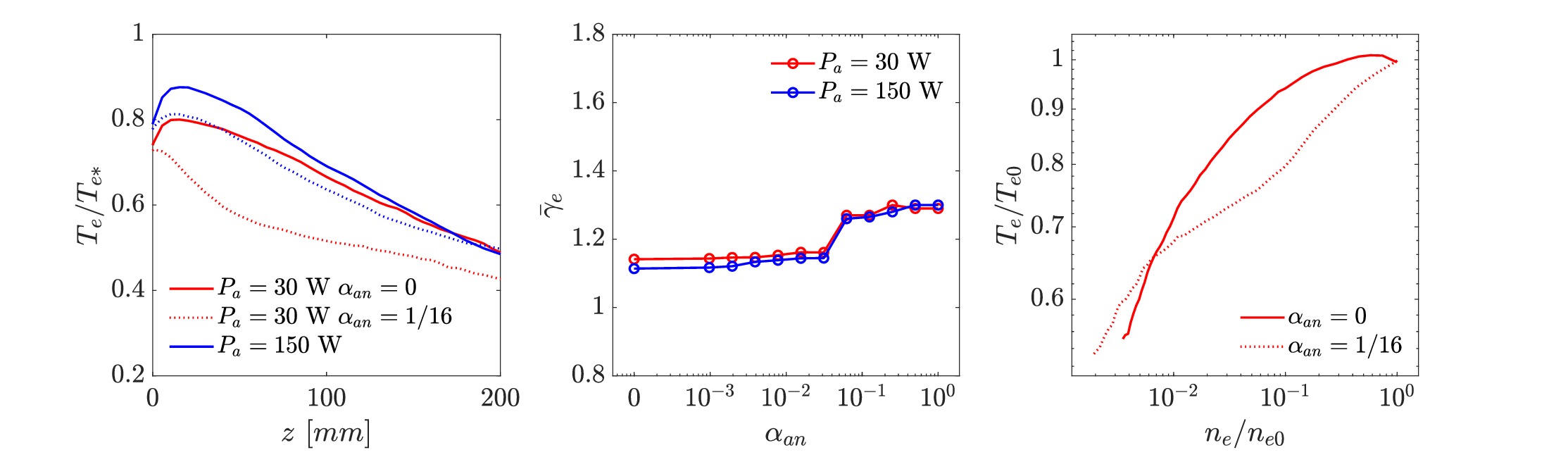}
 \caption{\label{fig:temp} (a) $T_e$ \textcolor{black}{on the symmetry axis for 30 and 150 W for $\alpha_{an}=0$ and 1/16; (b) the corresponding average polytropic indices for increasing $\alpha_{an}$; (c) polytropic fitting on the symmetry axis for $P_a=30$ W.}}
 \end{figure*}

\textcolor{black}{The anomalous collisions dominate the electron cooling for large values of $\alpha$ (e.g. $\alpha = 1/16$). \Fref{fig:temp}~(a) shows the effect of increasing the anomalous diffusion on the spatial profile of $T_e$. The slow classical cooling in the upstream region does not fit with experimental evidences. As $\alpha_{an}$ increases, the electron cooling also increases and lowers $T_e$ downstream, with a far more rapid decay in the upstream. It is key to recall that while $T_{e*}$ is fixed, and is the temperature of the forward-marching injected electrons only, the actual temperature at the inlet $T_{e0}$ lowers with increasing $\alpha_{an}$ because it also depends on the temperature of the reflected electron population returning to the source. As discussed previously, this depends highly on the magnitude of $\phi_\infty$, which determines the cut-off energy for those high energy electrons that will be missing from the electron population reflected back to the inlet. }

 \Fref{fig:temp}~(b) provides the corresponding polytropic indices for increasing anomalous collisionality. A transition occurs between values of 1.16 and 1.29 at $\alpha_{an}=1/32$. This confirms what is seen in the conversion efficiency; the plume is no longer magnetised for large values of $\alpha_{an}$. \Fref{fig:temp}~(c) shows the normalised values of $T_e$ versus $n_e$ in logarithmic scale along the axis. Many experimental and kinetic studies indicate a polytropic relation \textcolor{black}{\cite{b:ahedo2010,b:correyero2019,b:andrews2022,b:lafleur2014,b:little2016}},

 \begin{equation}
     \gamma_e = 1+\frac{ln{T_e/T_{e0}}}{ln{n_e/n_{e0}}}
 \end{equation}

 where $\gamma_e$ is the fitted polytropic index. Common values for xenon in these studies is 1.15-1.23 \textcolor{black}{\cite{b:collard2019,b:correyero2019,b:little2016,b:littlephd}}. With low $\alpha_{an}$ the expansion starts near-isothermal and approaches the unmagnetised value of 1.29 downstream once the electrons are detached from the MN, following the three region piecewise cooling regime observed in the previous study of similar MN expansions \cite{b:andrews2022}. The anomalous diffusion changes the cooling profile significantly. The initial expansion has a $\gamma_e \approx 1.25$ since the electrons are effectively unmagnetised due to the high combination of classical and anomalous collisions. Once the collisionality reduces, the electrons then become partially-magnetised again and $\gamma \approx 1.15$ before undergoing classical detachment, where $\gamma \approx 1.27$.

\section{Conclusion}
In conclusion, this study demonstrates the significant impact of anomalous diffusion on electron transport and the propulsive performance of magnetic nozzles in low-power cathode-less RF plasma thrusters through fully kinetic particle-in-cell (PIC) simulations. \textcolor{black}{This non-collisional transport mechanism, modelled in this work by a phenomenological Bohm-type anomalous collisionality, has been shown to substantially alter key plasma dynamics.} For the REGULUS-150-Xe thruster operating at low (30 W) and high (150 W) power conditions, enhanced cross-field electron transport inhibits the formation of the typical magnetic nozzle potential barrier, reducing radial confinement and lowering the downstream potential drop by up to 15\%. This suppression of steep pressure gradients diminishes diamagnetic currents, \textcolor{black}{and increases the paramagnetic \(E \times B\) current}. Consequently, magnetic nozzle efficiency drops from approximately 0.5 to 0.2, driven by reduced electron thermal energy conversion and increased plume divergence. At the Bohm limit of \(\omega_{ce}/16\), the simulation achieves agreement with experimental thrust profiles to within 20\%, a marked improvement over the 48\% overestimation observed in classical models at high power. These findings underscore the importance of accounting for anomalous transport effects in the accurate modelling of plasma thruster performance. Importantly, there exists a critical value of the Bohm coefficient where the MN expansion switches from a well-collimated to an under-collimated state as the electron transport transitions from that dominated by magnetic advection to that of cross-field diffusion. This critical value was found to be between 1/128 and 1/64.

\textcolor{black}{It is important to clarify that the MN efficiency of 0.2, for high $\alpha_{an}$, represents a worst-case scenario and is certainly not generally applicable. In reality, effective anomalous diffusion depends on factors such as thruster power (which dictates energy available to instability waves), mass flow rate, and other operational conditions. Consequently, a fully-turbulent Bohm approximation is unlikely to be the most accurate representation. As noted throughout this article, significantly lower values of the coefficient $\alpha_{an}$ have been found in other best-fit studies, underscoring the need for more predictive, self-consistent models beyond the Bohm approximation.}

\textcolor{black}{However, the impact of anomalous diffusion on MN efficiency remains a critical concern, given the strong dependence of performance on the consequences of enhanced cross-field transport. Takahashi proposed that instabilities in his thruster could induce inward transport, potentially improving MN performance \cite{takainward}. In previous work \cite{b:shauniepc}, it was theorised---using a drift-dissipative instability model---that this inward transport was likely due to the high-density conic structure specific to Takahashi’s thruster. Unfortunately, no other studies have confirmed inward transport, with most instead supporting outward diffusion, which is detrimental to performance. Encouragingly, the results from the drift instability model suggest that diffusion remains spatially confined to regions of high azimuthal electron drift \cite{b:shauniepc}, leading to significantly smaller performance losses compared to the Bohm diffusion presented here.}

\textcolor{black}{Additionally, it is important to acknowledge the strong coupling between the inductive source and the MN, a factor that cannot be fully captured in studies focusing solely on the MN. Although, Souhair et al. found minor differences in the plasma source discharge properties when Bohm scaling was increased \cite{b:souhairAIP}. At present, the primary performance limitation of plasma thrusters utilizing an MN remains in energy losses within the source tube during plasma production.}

\textcolor{black}{Future work will primarily focus on the proper use of anomalous transport in predictive simulations. This will require the further development of a self-consistent model of the anomalous diffusion as a function of the local plasma properties and allowing for behaviour varying across both power and mass flow rate regimes. The aforementioned preliminary work, taking the conjecture that anomalous transport in MNs arises from a kinetic drift-dissipative instability and deriving a self-consistent diffusion from wave quasi-linear theory, has yielded promising results in comparison to experiments \cite{b:shauniepc}.}

\appendix
\section{Global Model}
\label{sec:appendixA}
\setcounter{section}{1}

A 0D Global (volume-averaged) Model, presented previously in references \cite{b:souhairPP,b:AA2022}, was adopted to evaluate the plasma properties in the RF source discharge. The dynamics in the source are governed by the species density and energy conservation,
\begin{equation}
    \frac{d{n'_k}}{dt}=Y_{chem}^k-Y_{wall}^k-Y_{*}^k+Y_{in}^k
\end{equation}
\begin{equation}
    \frac{d}{dt}\left(\frac{3}{2}{n'}_e\bar{T}_e\right)=P_{a}-P_{chem}-P_{wall}-P_{*}
\end{equation}
where $n'_k$ is the number density of species $k$, including excited neutral states. $Y_{chem}^k$, $Y_{wall}^k$, $Y_{*}^k$ and $Y_{in}^k$ are the species $k$ density flux source/sink terms associated with plasma reactions, wall losses, particle outflow, and particle inflow respectively. $P_{a}$, $P_{chem}$, $P_{wall}$ and $P_{*}$ are the RF power coupled to the plasma, along with the source/sink terms associated with plasma reactions, wall losses, and particle outflow respectively. It should be noted that in these equations $n'_{e,i}$ is the peak plasma density under the RF antenna, whereas $n'_g$ is assumed constant according to the source pressure, and $\bar{T}_e$ is a volume-average.

The plasma reactions considered are electron elastic scattering, ionisation, and excitation. Therefore, $Y^k_{chem}$ and $P_{chem}$ are given as
\begin{equation}
    Y_{chem}^k = \sum_J K_{kJ}n'_Jn'_e-\sum_J K_{kJ}n'_kn'_e
\end{equation}
\begin{equation}
    P_{chem} = \sum_{k,J} K_{kJ}n'_kn'_e\Delta U_{kJ} + \sum_k K_{kk}n'_kn'_e\frac{3m_e}{m_k}\bar{T}_e
\end{equation}
where $K_{kJ}$ is the rate constant for the inelastic transitions from species $k$ to $J$, $K_{kk}$ is the rate constant for elastic collisions between species $k$ and electrons, and $\Delta U_{kJ}$ is the energy gap between species $k$ and $J$.

The Bohm sheath criterion is enforced at the source walls, and with a sonic source tube exit, similar expressions hold for $Y_{wall}^k$ and $Y_*^k$
\begin{equation}
    Y_{wall}^k=\frac{S^k_{wall}}{V}\Gamma^k_{wall}
\end{equation}
\begin{equation}
    Y_*^k=\frac{S_*^k}{V}\Gamma_*^k
\end{equation}
where $V$ is the volume of the source, $S^k$ is the equivalent surface area of the source, and $\Gamma^k$ is the particle flux. 
For ions and electrons $\Gamma_i=\Gamma_e = n'_eu_B$. Considering the equivalent surface area,
\begin{equation}
    S^e = S^i = h_L A_0 + h_R(A_R-A_c) + h_cA_c,
\end{equation}
where $h_L$, $h_R$ and $h_c$ are the semi-empirical axial, radial, and cusp edge-to-centre density ratios respectively. $A_R$ and $A_c$ are the lateral surface area and equivalent area of the cusp loss widths. A full description of these terms is given in reference \cite{b:AA2022}. For neutrals $Y^g_{wall}=-Y_{wall}^i$, assuming total recombination. The equivalent neutral surface is equal to the source tube exit cross-section, $S_*^g = A_0$ and, assuming the neutrals are in the free-molecular regime, $\Gamma^g_*=1/4n'_{g}\bar{u}_{g}$.\\
From the Bohm sheath criterion, the energy terms subsequently read
\begin{equation}
    P_{wall,*} = Y^e_{wall,*}\left( 2\bar{T}_e\frac{1}{1-\alpha}-2T_s\frac{\alpha}{1-\alpha}+\phi_{wall,*}\right)
\end{equation}
where $\alpha$ is the secondary electron emission coefficient, $T_s$ is the secondary electron emission temperature, and $\phi_{wall,*}$ is the sheath potential given by
\begin{equation}
    \phi_{wall,*} = \bar{T}_e\ln\left(\sqrt{\frac{m_i}{2\pi m_e}}(1-\alpha)(1-\delta) \right)
\end{equation}
where $\delta$ is the electron elastic reflection coefficient. In the case of the source tube exit, $\alpha=\delta=0$. For the walls, according to reference \cite{b:souhairPP}, $\delta=\delta_0 E_r^2/(\bar{T}_e+E_r)^2$ and $\alpha=2\bar{T}_e/E_s$. The following were assumed in this work: $T_s=2$~eV, $\delta_0=0.4$, $E_r=20$~eV, and $E_s=50$~eV.

Regarding the inflow, only neutral gas is injected into the source
\begin{equation}
    Y_{in}^g = \frac{\dot{m}}{Vm_g}
\end{equation}
where $\dot{m}$ is the propellant mass flow rate.

Given the thruster input power $P_{RF}$, propellant mass flow rate $\dot{m}$, and magnetic field $\mathbf{B}$, the Global Model solves the system of equations to provide $\dot{m}_{i*}$, $\dot{m}_{g*}$ and $T_{e*}$ to the PIC model,
\begin{eqnarray}
    &\dot{m}_{i*} = m_ih_Ln'_iu_BA_0 \\ 
    &\dot{m}_{g*} = m_g n'_g\bar{u}_gA_0 \\ 
    &T_{e*} = \bar{T}_e 
\end{eqnarray}

\section*{Acknowledgements}
This work was supported by Technology for Propulsion and Innovation (T4i) S.p.A., where REGULUS-150-Xe is currently being developed in the frame of the ESA Contract No.4000130900/20/NL/RA. Acknowledgement is also given to the CINECA award under the ISCRA initiative, for the availability of high-performance computing resources and support.

\section*{References}
\bibliographystyle{ieeetr}
\bibliography{bibliography}

\end{document}